\documentclass[twocolumn,tighten,twocolappendix]{aastex631}
\usepackage{amsmath}
\usepackage{enumerate}
\usepackage{array}
\usepackage[shortlabels]{enumitem}
\setlist[enumerate]{nosep,nolistsep}
\setlist[itemize]{nosep,leftmargin=*}

\RequirePackage{xcolor,colortbl}

\newcommand{\modest}{\mbox{MODEST}}

\definecolor{darkgreen}{rgb}{0.0, 0.4, 0.0}

\usepackage{txfonts}
\usepackage{booktabs}

\definecolor{al}{rgb}{0.6,0.2,0.0}
\definecolor{as}{rgb}{0.8,0.2,0.3}
\definecolor{mvn}{rgb}{0.5,0.5,0.9}

\newcommand{\sks}[1]{{\color{red}{}}}

\graphicspath{{Figures_BLB_common_appearance/}{Figures_BLB_common_appearance/Figures_Stokes_prof_strongB/}{Figures_BLB_common_appearance/Isolated_BLBs/}{Figures_BLB_common_appearance/Figure_stokes_profiles/}}
\usepackage{multirow}
\shorttitle{Superstrong magnetic fields in bipolar light bridges}
\shortauthors{{Castellanos~Dur\'{a}n} et al}

\begin{document}

\correspondingauthor{J.~S. {Castellanos~Dur\'an}}
\email{castellanos@mps.mpg.de}

\author[0000-0003-4319-2009]{J.~S. {Castellanos~Dur\'an}}
\affiliation{Max Planck Institute for Solar System Research, Justus-von-Liebig-Weg 3, D-37077 G\"ottingen, Germany}

\author[0000-0003-1459-7074]{A. Korpi-Lagg}
\affiliation{Max Planck Institute for Solar System Research, Justus-von-Liebig-Weg 3, D-37077 G\"ottingen, Germany}
\affiliation{Department of Computer Science, Aalto University, PO Box 15400, FI-00076 Aalto, Finland}

\author[0000-0002-3418-8449]{S.~K. Solanki}
\affiliation{Max Planck Institute for Solar System Research, Justus-von-Liebig-Weg 3, D-37077 G\"ottingen, Germany}

\author{M. van~Noort}
\affiliation{Max Planck Institute for Solar System Research, Justus-von-Liebig-Weg 3, D-37077 G\"ottingen, Germany}

\author[0000-0001-9166-8230]{N. Milanovic}
\affiliation{Max Planck Institute for Solar System Research, Justus-von-Liebig-Weg 3, D-37077 G\"ottingen, Germany}

\newcommand{\Numquietsun}{5.0$\times10^{7}$}       
\newcommand{\Numpenumbra}{1.8$\times10^{7}$}       
\newcommand{\Numwarmumbra}{3.9$\times10^{6}$}      
\newcommand{\Numcoldumbra}{5.0$\times10^{5}$}      
\newcommand{\NumTOTALumbra}{4.4$\times10^{6}$}     
\newcommand{\Numbadpixels}{53\,000}                

\newcommand{\Percquietsun}{68.6\,\%}               
\newcommand{\Percpenumbra}{25.3\,\%}               
\newcommand{\Percwarmumbra}{5.3\,\%}               
\newcommand{\Perccoldumbra}{0.7\,\%}               
\newcommand{\Percbadpixels}{0.07\,\%}              

\newcommand{\PercColdumbraWRTtotalumbra}{11.4\,\%}
\newcommand{\PercTOTALumbra}{6.0\,\%}              

\newcommand{\numtiles}{22\,000}                    
\newcommand{\numinvs}{869}                         
\newcommand{\numars}{78}                           
\newcommand{\numpixels}{7.3$\times10^{7}$}         
\newcommand{\numdatapoints}{3.2$\times10^{10}$}    

\newcommand{\numblb}{98}   
\newcommand{\numblbAR}{51} 
\newcommand{\numblbscans}{448}

\newcommand{\numblbsightings}{630} 

\newcommand{\highestB}{0.95\,Tesla}
\newcommand{\ARhighestB}{AR\,12345}
\newcommand{\datehighestB}{2015, July 4}

\newcommand{\numBLBobservedmorethanonce}{66.3\%}
\newcommand{\numBLBobservedmorethanonceratio}{66.3\%\,(65/98)}


\newcommand{\areablb}{\mbox{35.4$^{+30.6}_{-17.8}$}\,Mm$^2$}
\newcommand{\widthblb}{\mbox{3.1$^{+2.0}_{-0.9}$}\,Mm}
\newcommand{\lengthblb}{\mbox{14.0$^{+7.4}_{-5.4}$}\,Mm}

\newcommand{\aspectratioblb}{1:18.7}

\title{Superstrong magnetic fields in sunspot bipolar light bridges}

\begin{abstract}
Recent solar observations of bipolar light bridges (BLBs) in sunspots have, in a few individual cases, revealed magnetic fields up to 8.2\,kG, which is at least twice as strong as typical values measured in sunspot umbrae. However, the small number of such observations hinted that such strong fields in these bright photospheric features that separate two opposite-polarity umbrae, are a rare phenomenon. We determine the field strength in a large sample of BLBs with the aim of establishing how prevalent  such strong fields are in BLBs. We apply a state-of-the-art inversion technique that accounts for the degradation of the data by the intrinsic point spread function of the telescope, to the so far largest set of spectropolarimetric observations, by Hinode/Solar Optical Telescope spectropolarimeter, of sunspots containing BLBs. We identified 98 individual BLBs within 51 distinct sunspot groups. Since 66.3\% of the BLBs were observed multiple times, a total of 630 spectropolarimetric scans of these 98 BLBs were analysed. All analysed BLBs contain magnetic fields stronger than 4.5\,kG at unit optical depth. The field strengths decrease faster with height than the fields in umbrae and penumbrae. BLBs display a unique continuum intensity and field strength combination, forming a population well separated from umbrae and the penumbrae. The high brightness of BLBs in spite of their very strong magnetic fields points to the presence of a so far largely unexplored regime of magnetoconvection. 
\end{abstract}
\vspace{-1cm} 

\section{Introduction}

Sunspot structure and evolution is controlled by the magnetic field and its flows. In the umbra, vertical fields in the range of 3\,kG or higher are usually found, while in the penumbra, inclined fields typically range between 1--2.5\,kG \citep[e.g.,][]{Solanki2003, Borrero2011LRSP, Tiwari2015A&A...penumbra}.
Larger sunspots usually have a darker umbra, implying the presence of  stronger magnetic fields \citep[e.g.,][]{Mathew2007A&A...spotcontrast, Schad2010SoPh..UmbralB, Schad2014SoPh..UmbralB, Kiess2014A&A...Bumbra}. However, the umbral darkness (i.e., low temperature) makes umbral magnetic field measurements extremely difficult: the photon statistics is low, the effect of stray light from the surrounding, more than 10 times brighter quiet-sun regions becomes stronger, and in many spectral regions the appearance of molecular lines complicates the identification of the Zeeman components of atomic lines. Moreover, extremely large sunspot groups do not appear often in the Sun. Despite these constraints, large fields have been observed in sunspot umbrae with values ranging from 3\,kG to up to $\sim$6\,kG \citep[e.g.,][]{Livingston2006SoPh, Pevtsov2014SoPh..CyclicB}, but it is probably safe to say that some uncertainty remains regarding the real value of the strongest magnetic field in large and cool sunspot umbrae. For medium to small sunspots the typical umbral magnetic fields are in the $2.5$--3.5 kG range \citep[e.g,][]{MartinezPillet1993A&A...B2cont, Collados1994A&A...spots, Mathew2003A&A...spot, Pevtsov2011ApJ...LongtermB, Jurcak2018A&A...JurcakCriterion,Li2022ApJ...I2Binsunspots}.

In the past two decades, observers and modellers have documented magnetic fields in sunspots exceeding typical umbral values. Locations of such ``superstrong" fields are the outer ends of penumbral filaments \citep{vanNoort2012A&A}, in counter Evershed flows \citep{Siu-Tapia2017A&A, Siutapia2019, CastellanosDuran2023...ejectionCEFs}, or on so-called bipolar light bridges (BLBs) \citep[e.g.][]{Tanaka1991SoPh, Zirin1993b, Livingston2006SoPh, Okamoto2018ApJ, Wang2018RNAAS, Toriumi2019ApJL...strongB, CastellanosDuran2020, Hotta2020MNRAS...strongB, Lozitsky2022ApJ...Strongfields,Schuck2022ApJ...strongfields, Liu2023ApJ...strongfieldandflows}.

The term `superstrong fields' indicates that the field strengths are significantly stronger than usual in a given type of solar feature. Specifically for the purposes of this Letter, we compare the field strengths to typical values in the umbra and refer to strong magnetic fields as those between 3-5 kG, and superstrong magnetic fields as those stronger than 5 kG, following \citet{CastellanosDuran2022...Phd}.

BLBs appear as bright regions between two umbrae with opposite magnetic polarities. The polarity inversion line is generally located along the long axis of a BLB, and therefore $\delta$-sunspot groups are the only places where these features exist. There is some evidence of BLBs being formed by the coalescence of sunspots of opposite polarity \citep[e.g.,][]{ZirinWang1993Natur...BipolarLightBridge}, but the mechanism by which they form remains unclear.

To our knowledge, there is only one statistical study of sunspots with strong magnetic fields that uses archival data, by \citet{Livingston2006SoPh}. This study included 55 ARs with magnetic fields over 4\,kG, five had magnetic fields over 5\,kG, and one active region appeared on February 28, 1942 (AR\,7378) with a maximum field 
strength of 6.1\,kG \citep{Livingston2006SoPh}. However, \citet[][]{Livingston2006SoPh}'s archive study relied on spectroscopic data at low spatial and spectral resolution and focused mainly on umbrae. In the present Letter, we present another such statistical study, but this time of the strongest magnetic field in BLBs using spectropolarimetric observations and state-of-the-art inversions of the radiative transfer equation.

One of the  first clear detections of strong magnetic fields in BLBs was reported by \citet{Okamoto2018ApJ}, who measured 6.2 kG using the Milne-Eddington approximation to model the observations. However, due to gradients in the magnetic field and line-of-sight velocities with height, this BLB exhibits complex Stokes profiles.
\citet{CastellanosDuran2020} reanalysed the same observations with the help of advanced 2D coupled inversions \citep{vanNoort2012A&A, vanNoort2013A&A} that account for the smearing of the information by the telescope’s point-spread function (PSF) and height gradients of the atmospheric parameters. These authors have not only confirmed the presence of superstrong fields, but they found an even larger maximum value of 8.2 kG in the deepest layer sampled by the \ion{Fe}{1} 630\,nm line pair \citep{CastellanosDuran2020}.

Recent MHD simulations of $\delta$-sunspot groups also found magnetic fields exceeding 6\,kG in a highly sheared BLB \citep{Toriumi2019ApJL...strongB, Hotta2020MNRAS...strongB}.

All the recent studies about strong magnetic fields focused on AR\,11967 and AR\,12673 \citep[e.g.,][]{Okamoto2018ApJ, Wang2018RNAAS,Lozitsky2022ApJ...Strongfields,Schuck2022ApJ...strongfields, Liu2023ApJ...strongfieldandflows}. 
The presence of strong fields was clearly established in these publications for these regions, with some authors suggesting that these two regions were ``special" in some way \citep[e.g.,][]{Sun2017RNAAS...strongfields}. Both ARs are among the largest during solar cycle 23, and AR\,12673 hosted the largest X-class flare of the cycle \citep[e.g.,][]{Yang2017ApJ...XflareBLB, Verma2018A&A}. The current work presents evidence that the appearance of strong to superstrong magnetic fields is common in ARs with $\delta$-spots, as all BLBs in such spots (in our sample) contain strong fields and a very large fraction contain superstrong ones.

\begin{figure}[htb]
\includegraphics[width=.48\textwidth]{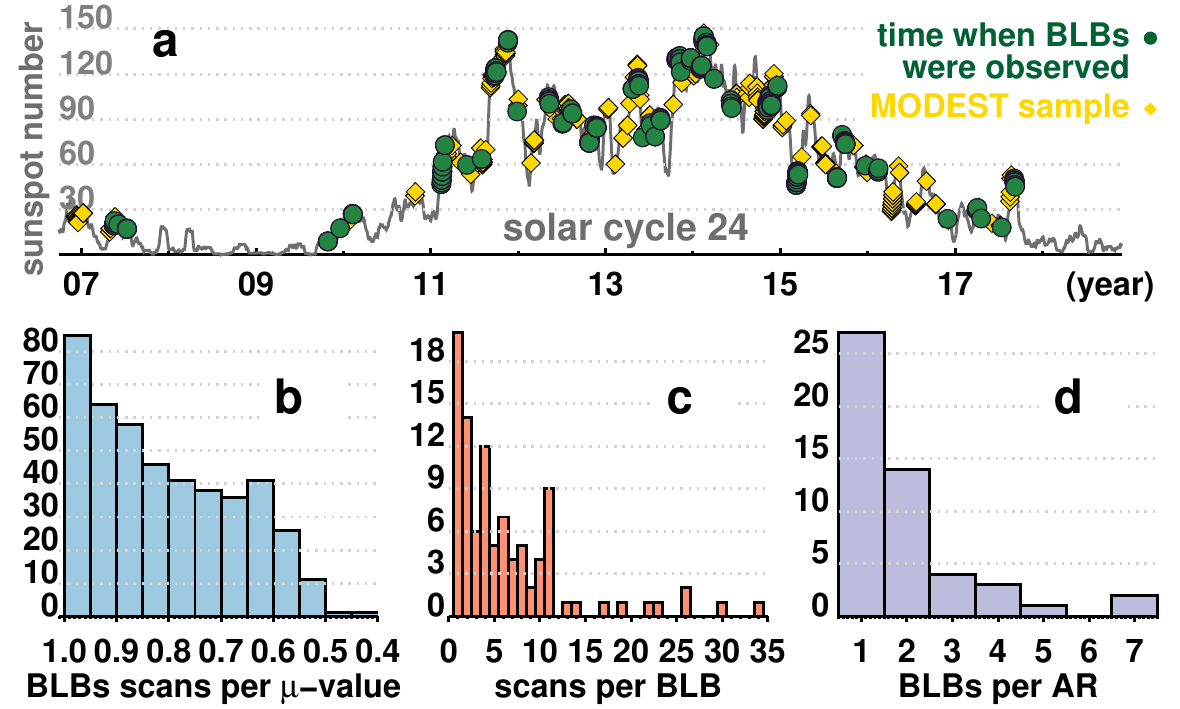}
\centering
 \caption{Characteristics of Hinode/SOT-SP scans harbouring BLBs. Panel (a) shows the smoothed daily sunspot number (grey line). Yellow diamonds mark all scans as part of the MODEST sample. Green circles indicate the observation days when BLBs were found. Histograms show the number of Hinode/SP scans of BLBs within a given $\mu$--range (b), the number of times a BLB was scanned (c), and the number of BLBs per sunspot group (d). }\label{fig:arsample}
 \end{figure}

BLBs have an intensity similar to penumbrae, with  temperatures higher than the dissociation temperature of typical molecules found in sunspots. Therefore, observations of these regions are not affected by molecular blends. Also, observations of BLBs have a larger signal-to-noise ratio than those obtained in umbrae. These two facts make the results of inversions of the spectropolarimetric data more reliable. In this study, we aim to establish how rare or common the superstrong fields observed in BLBs are, such as those found in AR\,11967 or AR\,12673, are. We search for such fields in the archive of the Solar Optical Telescope Spectropolarimeter  \citep[SOT-SP;][]{Ichimoto2008SoPh,Tsuneta2008,Lites2013SoPh} onboard Hinode \citep{Kosugi2007}. Specifically, we used the \modest{} catalogue, which covered 13\,yr at the time this work started of Hinode observations starting in 2006 \citep[see][for further details]{CastellanosDuran2024...modest}.

\section{Observations and data processing}

Spectropolarimetric inversions of ARs with sunspots are taken from the \modest{} catalogue. When we started working on BLBs, the catalogue comprised \numinvs{} inverted spectropolarimetric scans of \numars{} ARs taken by Hinode/SOT-SP. Data were recorded between 2006 December 8 and 2019 July 27. Yellow diamonds in Fig.~\ref{fig:arsample}a show the time stamps of the \modest{} inversions, starting at the end of solar cycle 23 and covering most of cycle 24. \modest{} uses spatial scans produced by SOT-SP onboard Hinode. The \modest{} sample contains a disproportionately high number of $\delta$-sunspot groups \citep{CastellanosDuran2024...modest}, making it ideal for the present study. Hinode/SOT-SP records the full Stokes spectrum $(I(\lambda), Q(\lambda), U(\lambda), V(\lambda))$ at a spectral sampling of 21.5\,m\AA{} of the \ion{Fe}{1} line pair at $6301.5$\,\AA \ and $6302.5$\,\AA, whose Land\'e factors are 1.67 and 2.5, respectively.

 \begin{figure*}[htbp]
     \centering
\includegraphics[width=.795\textwidth]{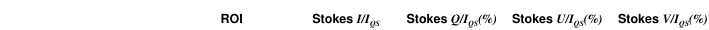}
\includegraphics[width=.795\textwidth]{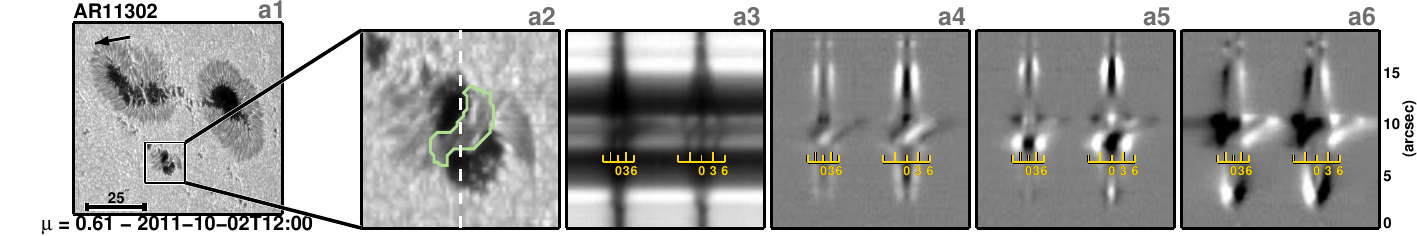} 
\includegraphics[width=.795\textwidth]{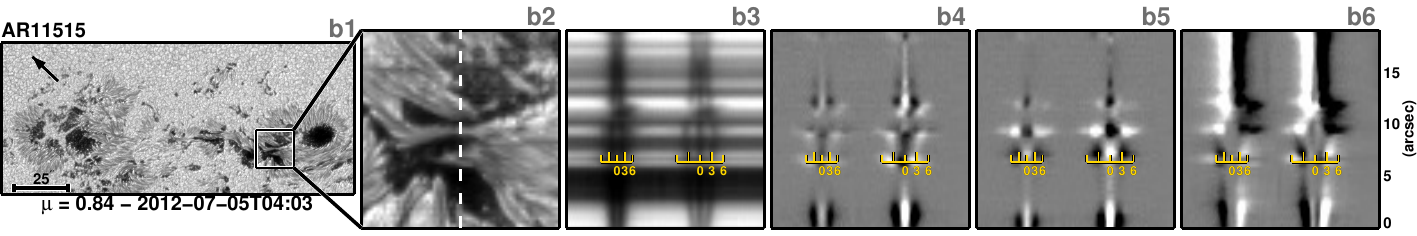} 
\includegraphics[width=.795\textwidth]{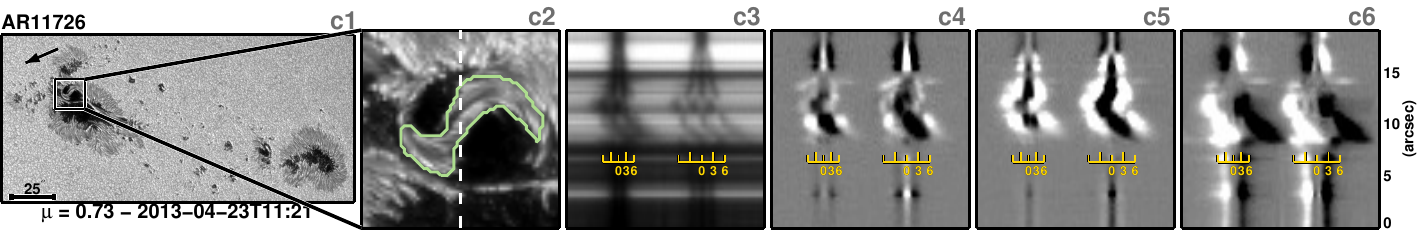}
\includegraphics[width=.795\textwidth]{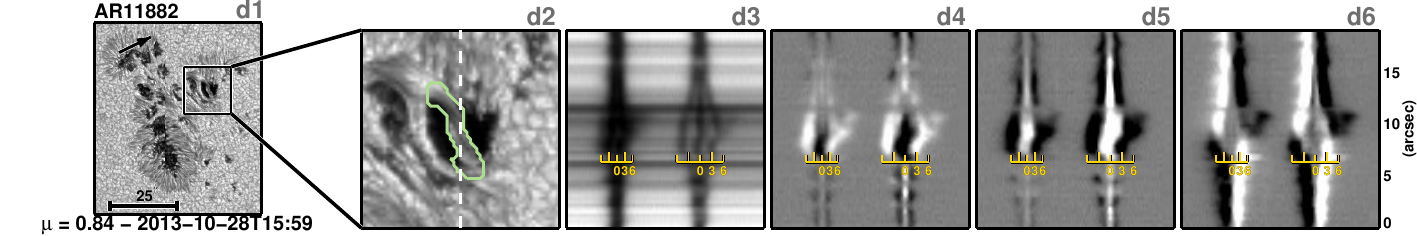} 
\includegraphics[width=.795\textwidth]{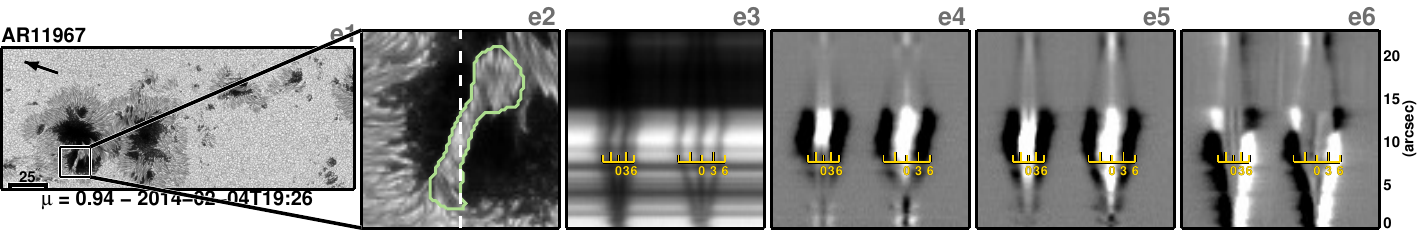} 
\includegraphics[width=.795\textwidth]{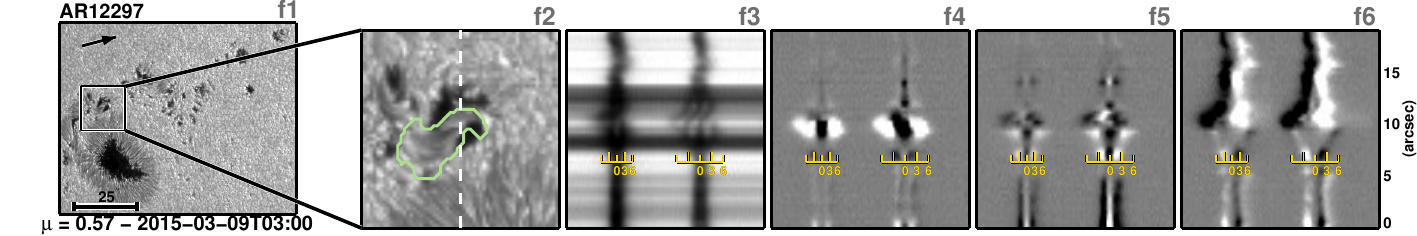} 
\includegraphics[width=.795\textwidth]{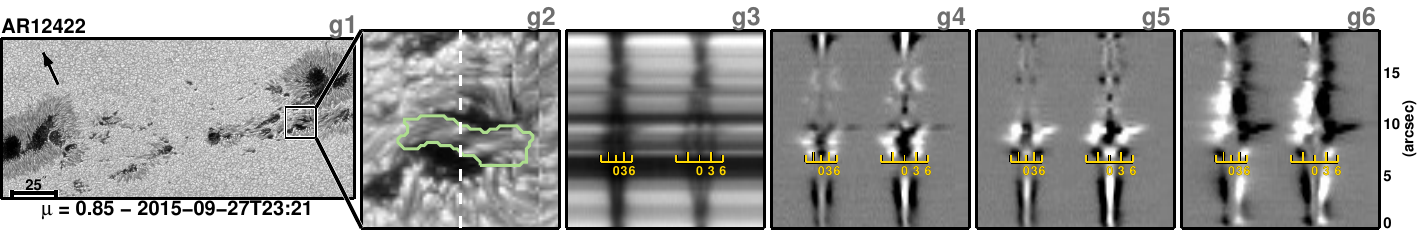} 
\includegraphics[width=.795\textwidth]{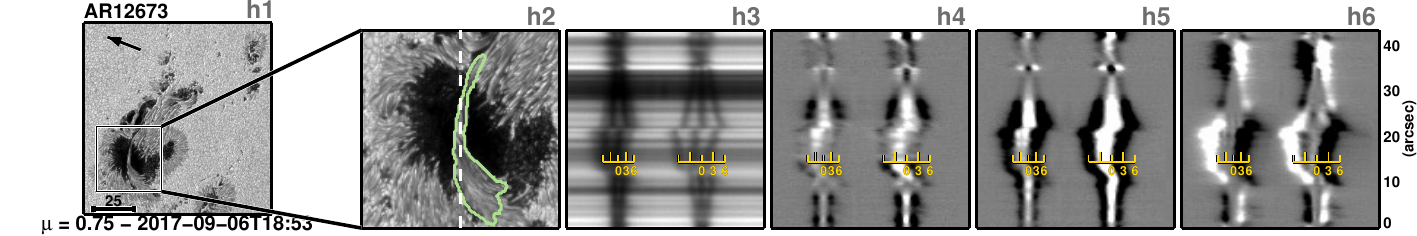}
\includegraphics[width=.795\textwidth]{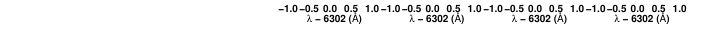}
     \caption{Bipolar light bridges and observed Stokes profiles along a cut through the BLBs. The left two columns show the continuum images of the AR and a zoom-in into the BLBs, outlined by the green contours. Arrows point towards disk centre. The $\mu$-value and time is given for the centre of each scan. The four columns on the right show the Stokes $I$, $Q$, $U$, and $V$ spectra, respectively, along the dashed line in column 2. The Stokes profiles are saturated at $\pm$5\%. The yellow ticks show the Zeeman splitting ($\Delta \lambda_{\rm Zeeman}=4.67 \times 10^{-13} g_{\text {eff }} \lambda_0B$) for a magnetic field of 0, 3, and 6\,kG.  }\label{fig:ObservedStokes}
 \end{figure*}

 \begin{figure*}[htbp]
     \centering
\includegraphics[width=.82\textwidth]{title_stokeprofiles2stokes.pdf}
\includegraphics[width=.82\textwidth]{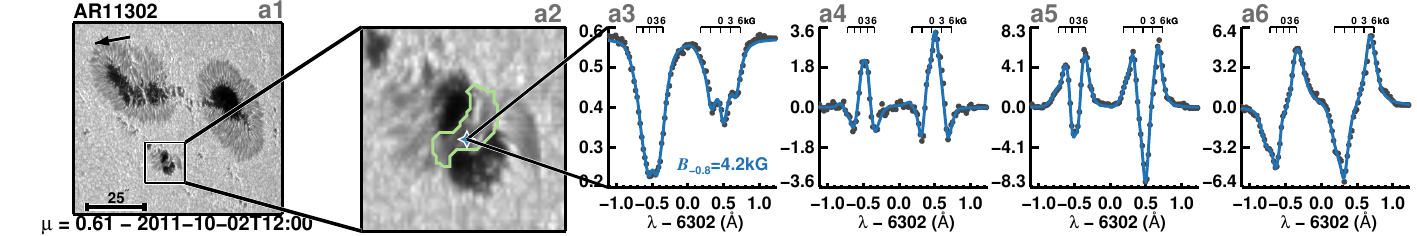} 
\includegraphics[width=.82\textwidth]{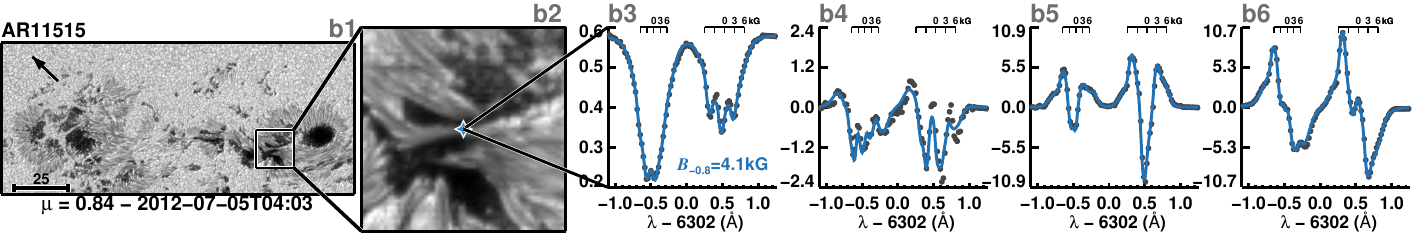} 
\includegraphics[width=.82\textwidth]{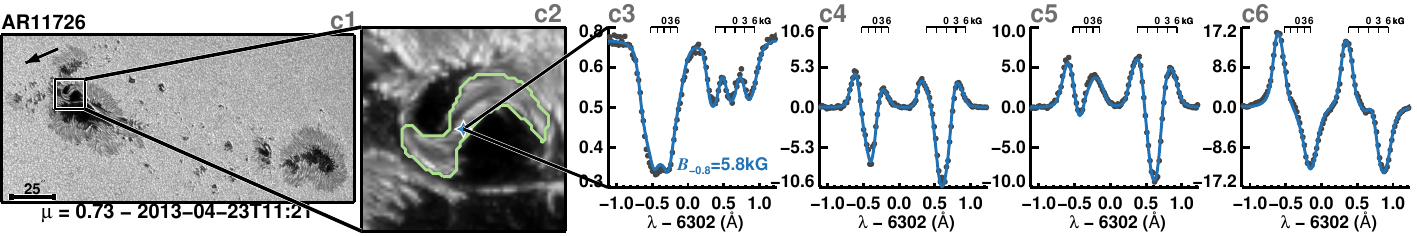}
\includegraphics[width=.82\textwidth]{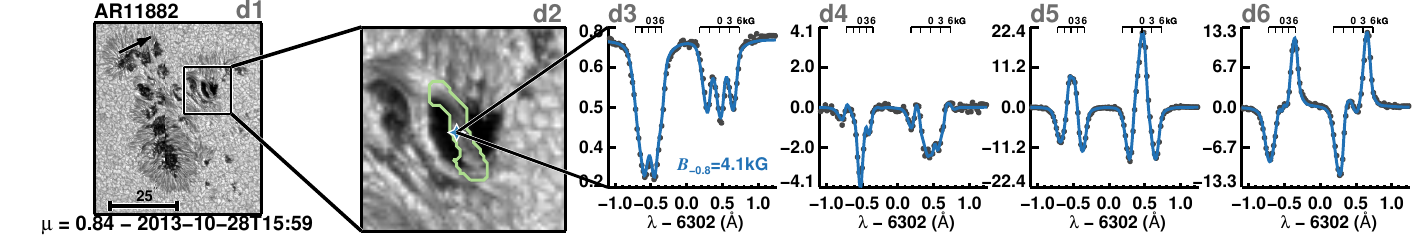} 
\includegraphics[width=.82\textwidth]{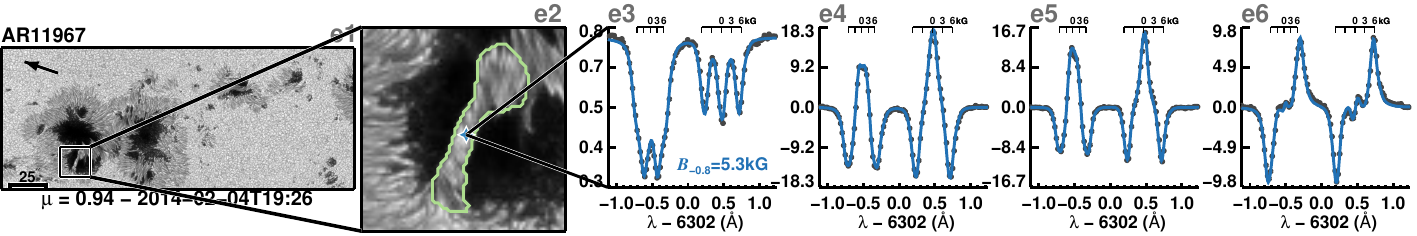} 
\includegraphics[width=.82\textwidth]{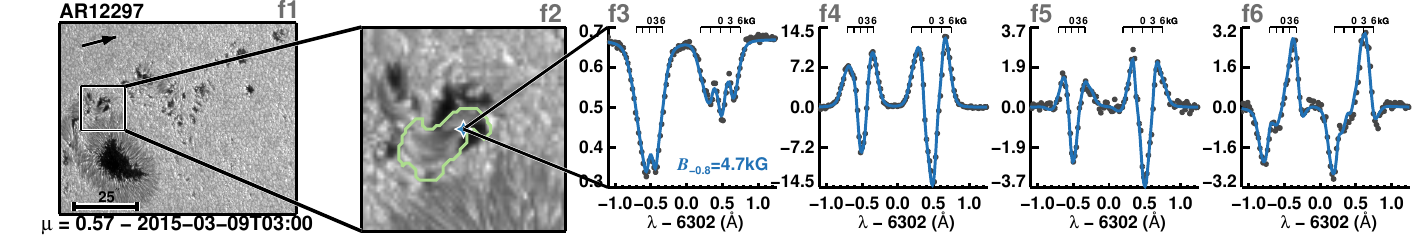} 
\includegraphics[width=.82\textwidth]{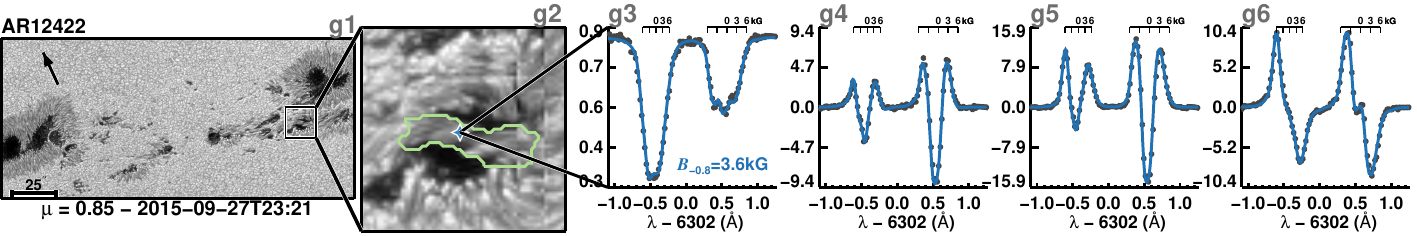} 
\includegraphics[width=.82\textwidth]{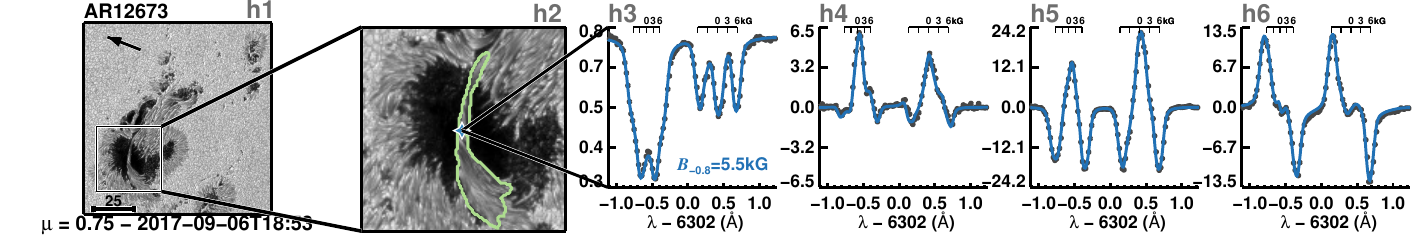} 

     \caption{Observed Stokes profiles at selected positions (blue star in the ROI column) within the BLBs and their MODEST fits. The left two columns show the continuum images of the AR and a zoom-in into the BLBs (green contours). Arrows point towards disk centre. The four columns on the right show the observed Stokes $I$, $Q$, $U$, and $V$ (gray dots). The blue lines show the best fit obtained with the spatially coupled inversion. The ticks on the top show the Zeeman splitting for a magnetic field of 0, 3, and 6\,kG. The wavelength position of the ticks take into account Dopplershifts at $\log_{10}\tau=-0.8$. The magnetic field strength at the same node is written in the third column. } \label{fig:FittedStokes}
 \end{figure*}

The \modest{} inversions were performed  using the spatially coupled version of the SPINOR code \citep{Solanki1987PhDT, Frutiger2000, vanNoort2012A&A, vanNoort2013A&A}. SPINOR assumes local thermodynamic equilibrium (LTE) and accounts for spatial smearing due to the telescope PSF. Atmospheric parameters vary with height: we set three optical depth nodes at $\log_{10}\tau = (-2.0,-0.8, 0)$ for the temperature, magnetic field strength, inclination and azimuth, and line-of-sight velocity. Microturbulence is assumed to be height independent. For further details on the \modest{} catalogue, see \citet{CastellanosDuran2024...modest}.

\begin{figure*}[htbp]
     \centering

    \setlength{\fboxrule}{1.6pt}
    \setlength{\fboxsep}{.5pt}
     \fbox{%
      \parbox{.99\textwidth}{
      \centering
      
    \includegraphics[width=.135\textwidth]{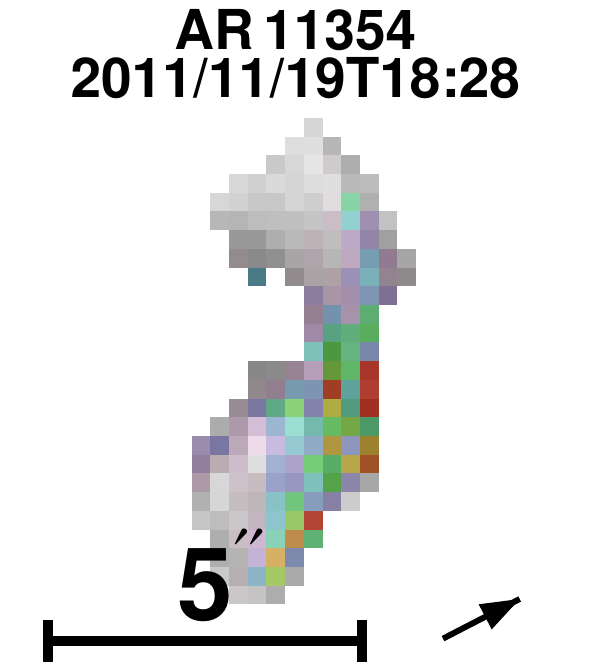}
    \includegraphics[width=.135\textwidth]{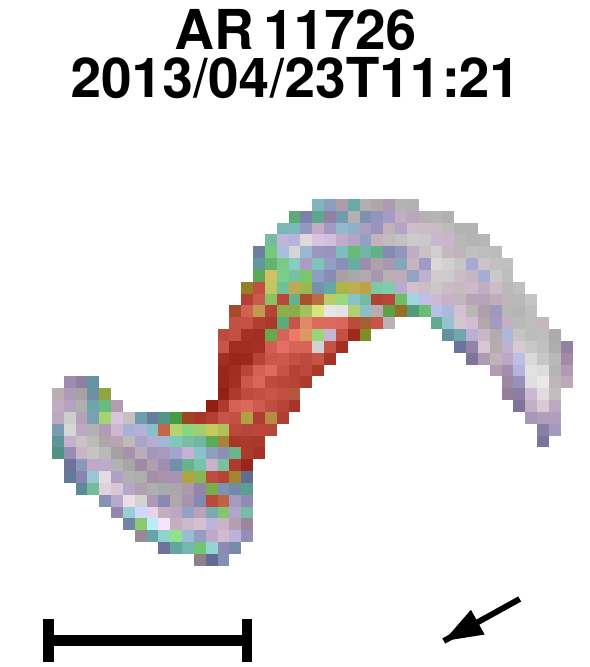}
    \includegraphics[width=.135\textwidth]{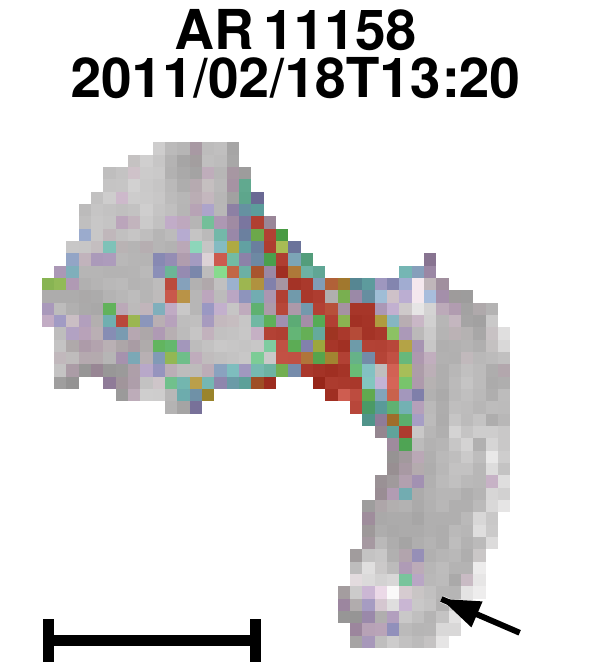}
    \includegraphics[width=.135\textwidth]{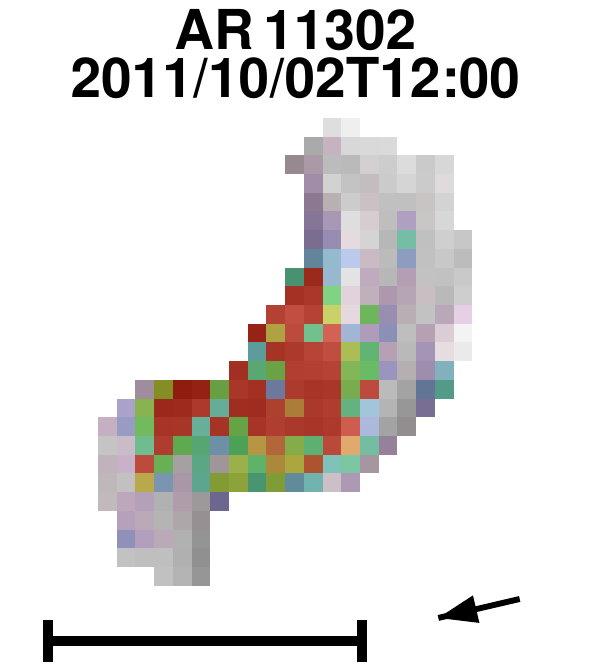}
    \includegraphics[width=.135\textwidth]{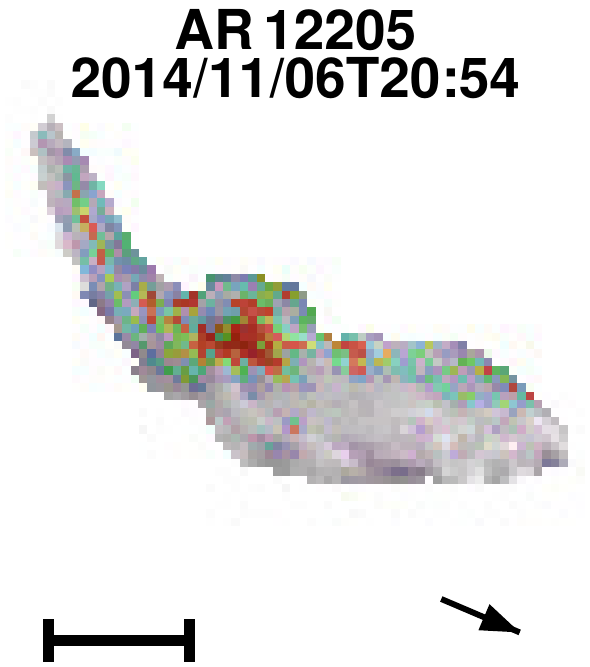}
    \includegraphics[width=.135\textwidth]{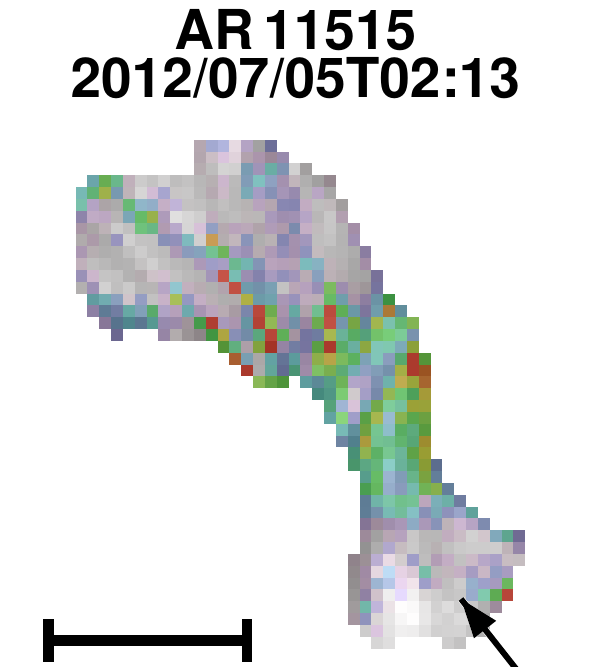}
    \includegraphics[width=.135\textwidth]{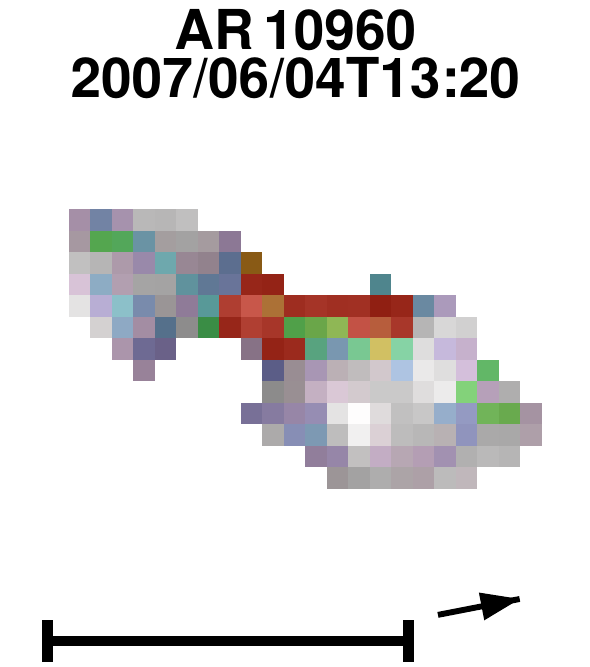}
    
    \includegraphics[width=.135\textwidth]{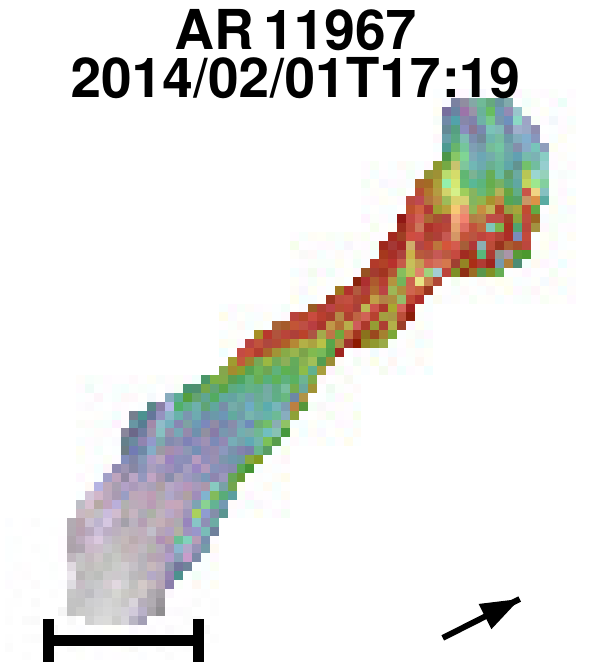}
     \includegraphics[width=.135\textwidth]{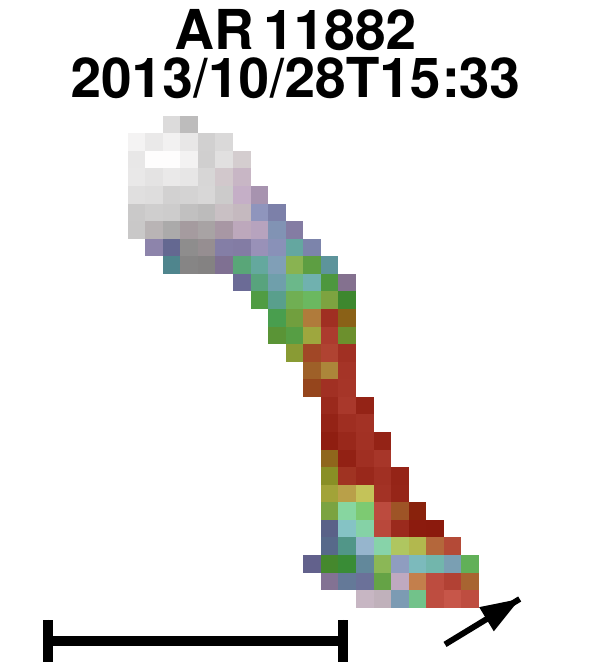}
    \includegraphics[width=.135\textwidth]{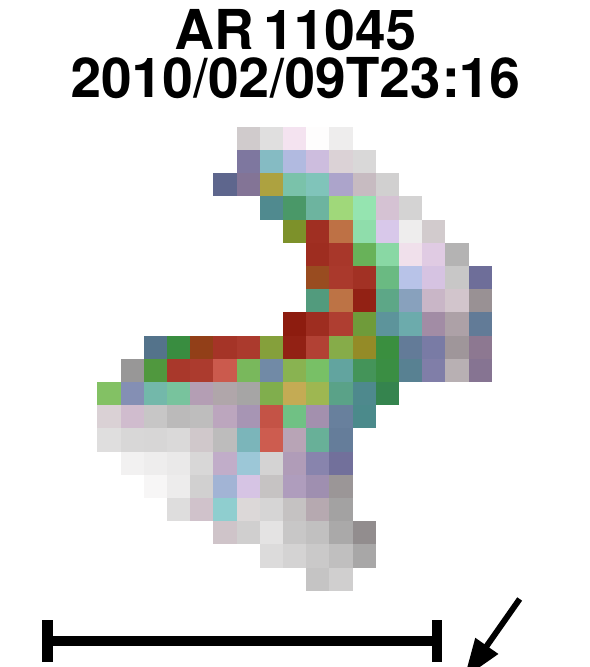}
     \includegraphics[width=.135\textwidth]{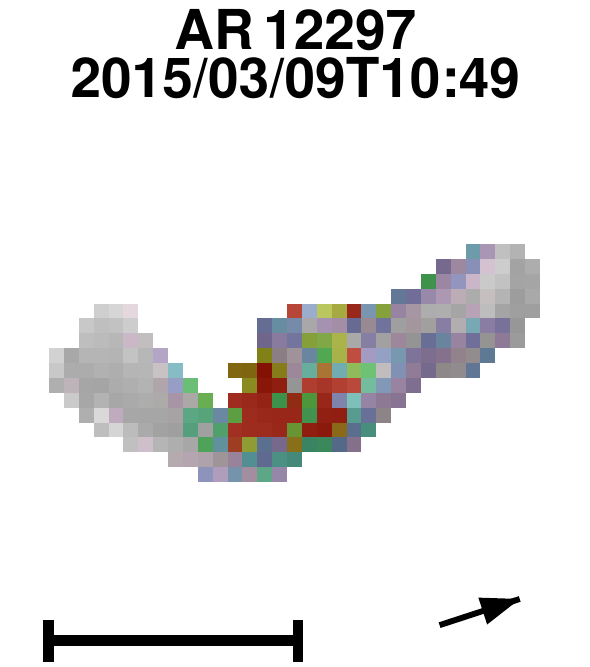}
    \includegraphics[width=.135\textwidth]{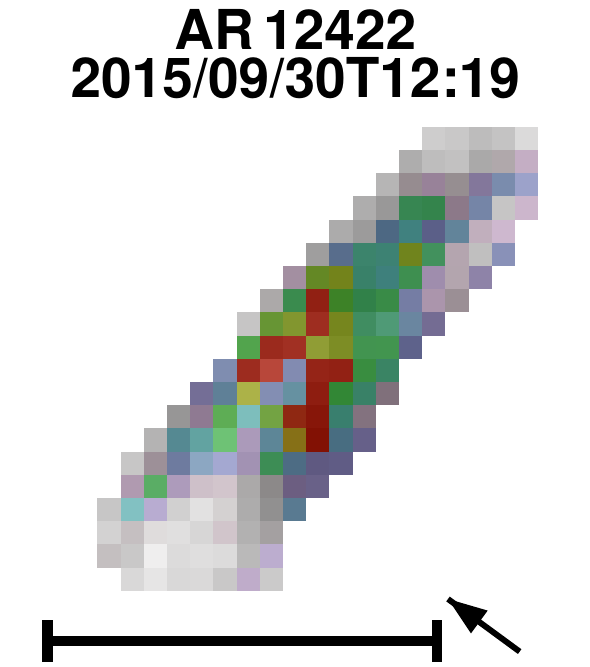}
    \includegraphics[width=.135\textwidth]{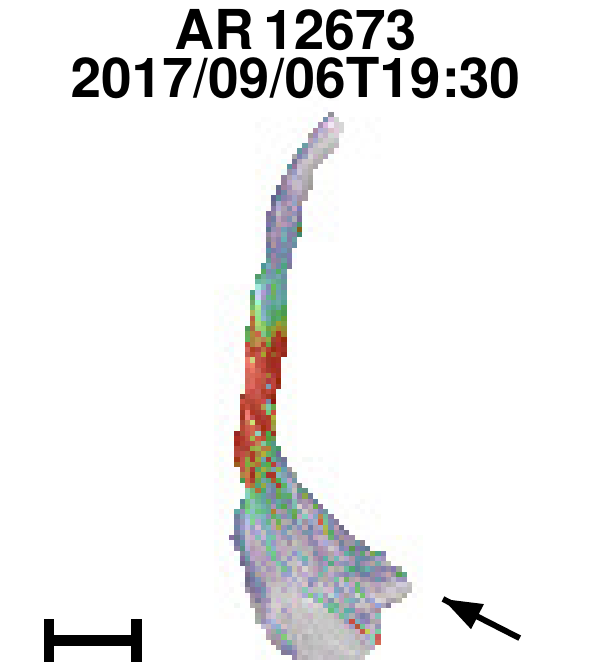}
    \includegraphics[width=.135\textwidth]{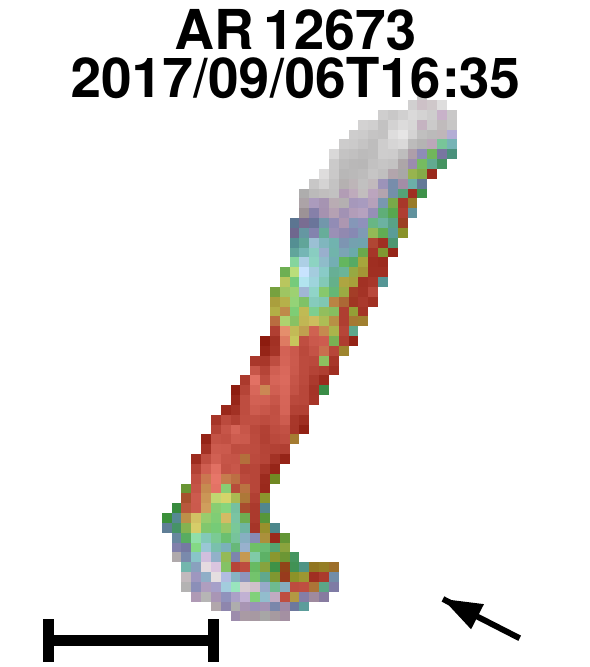}

    \includegraphics[width=.9\textwidth]{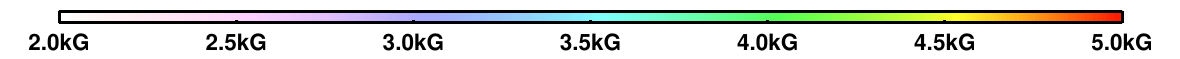}
     }}
     
     \caption{Continuum images of isolated BLBs. They are not plotted on the same scale, with the bars below the individual BLBs denoting 5\arcsec{}. The locations of strong magnetic fields within the BLB are colour shaded, with the shading saturated following the colour bar. The AR number and the scan time are written above each BLB. Arrows point towards disk centre.} 
     \label{fig:bones}
 \end{figure*}

Recent investigations have highlighted the importance of considering non-LTE (NLTE) effects on the formation of solar iron lines \citep{Smitha2020A&A......NLTEFeI, Smitha2021A&A...NLTEFeI, Smitha2022....NLTE6173}. The synthetic Stokes profiles showed significant changes when treated in NLTE. Nonetheless, for the simulation investigated by \cite{Smitha2020A&A......NLTEFeI}, where the errors in atmospheric parameters produced by neglecting NLTE effects were estimated, NLTE induced errors in the field strength amounted to values below 30\,G for sufficiently strong fields ($B>1$\,kG). Supplementary work is still needed to fully assess how NLTE effects impact the retrieved atmospheric conditions in the low solar atmosphere, but the work of \cite{Smitha2020A&A......NLTEFeI} suggests that departures from LTE are unlikely to influence our main results or conclusions.

\section{Results}

We systematically searched for BLBs in the \modest{} catalogue and identified \numblb{} separate BLBs as part of \numblbAR{} individual sunspot groups. Once a BLB was identified, we searched for all Hinode/SOT-SP scans that observed the same feature, independently of its evolutionary state. A total of \numblbscans{} Hinode/SOT-SP scans that contained BLBs were identified and analysed. A total of \numBLBobservedmorethanonceratio{} of BLBs were observed on more than one scan. In addition, in some cases multiple BLBs could be traced in a single Hinode/SOT-SP scan, depending on the complexity and size of the AR and the solar area covered by the scan. This led to \numblbsightings{} BLB scans, of which many were multiple observations of the same BLBs.

\begin{figure*}[htbp]
\centering
\includegraphics[width=1.\textwidth]{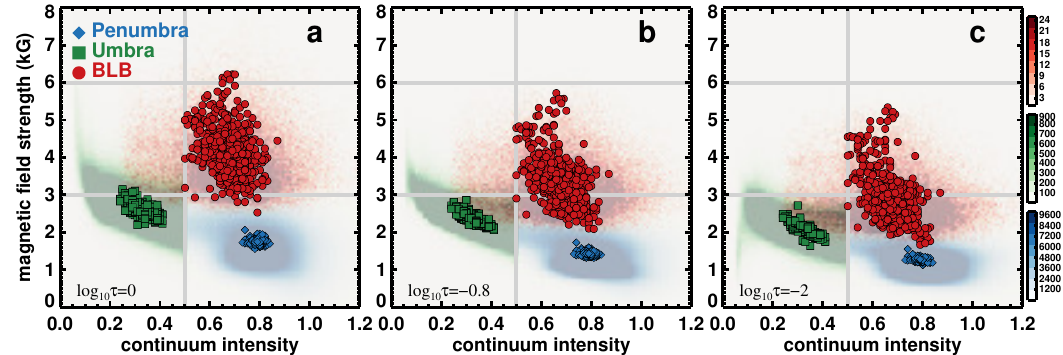}
    
\caption{Magnetic field strength at three optical depths, as marked in panels (a)-(c) as a function of continuum intensity. Colours refer to penumbrae (blue), umbrae (green), and BLBs (red). The shaded regions are composed of all the pixels in a given region for all Hinode/SOT-SP scans with BLBs. Each scan of a sunspot group with a BLB is represented by a symbol displaying the mean value of the 25\%\ strongest pixels within each of the three regions. As the number of pixels (area) covered by penumbra, umbra, and BLBs is comparatively different ($A_{\rm penumbra}>A_{\rm umbra}>A_{
\rm BLB}$), the background shades were saturated at different levels, as shown by the colour bars on the right, where the numbers represent how many pixels belong to a given bin.}
     \label{fig:AtmosphericConditionsBLBs}
 \end{figure*}

Figure~\ref{fig:arsample} summarizes Hinode/SOT-SP scans where BLBs were found regardless of their magnetic configuration. BLBs with strong magnetic fields were observed during all phases of the solar cycle. This is similar to the finding of \citet{Livingston2006SoPh} that sunspot umbrae with strong fields were found at all activity cycle phases. Figure~\ref{fig:arsample}b indicates that BLBs were more commonly seen at larger $\mu$--values, where $\mu$ is the cosine of the heliocentric angle. The $\mu$--values range from 1 to 0.41, and scans containing BLBs with $\mu$-values greater than 0.8 accounted for 56.5\%\,(253/448). This distribution results from the observation strategy. Hinode/SOT-SP mainly observes sunspots near the disk center to reduce foreshortening and other undesired effects at the limb. Figure~\ref{fig:arsample}c shows that most BLBs were found in multiple Hinode/SOT-SP scans, with a few having been scanned many times.

$\delta$-sunspot groups with two adjacent opposite-polarity umbrae within the same penumbra may host at least one BLB. In addition, over the evolution of an AR, multiple BLBs might appear (and disappear again) within it. Out of the 51 ARs, 24 of them (47\%) host multiple BLBs, while only 10 of them (19.6\%) have more than two BLBs (see Figure \ref{fig:arsample}d). In this respect the record holders are AR\,11515 and AR\,12297 where we counted seven BLBs each throughout their passage on the solar disk (see Appendix~\ref{sec:AR_many_BLBs}). Also, some BLBs split during their evolution into two fragments.  As a result, in these cases, we counted as one BLB the structure prior to fragmentation and as two new BLBs the structures after fragmentation. Similarly, when two BLBs merged into one, each was counted individually. In both cases, merging or fragmentation, the opposite-polarity umbrae remain separated by a BLB (see Appendix~\ref{sec:merging_fragmentation_BLBs}). It is important to note that the \modest{} sample of ARs was selected mainly to cover the most complex ARs. However, even in this sample, 34.6\% (27/\numars{}) of ARs do not contain BLBs (compare green and yellow symbols in Fig.~\ref{fig:arsample}a).

\begin{table}[htbp]
\centering
\addtolength{\tabcolsep}{-4pt}
\begin{tabular}{cccc}
\toprule[1.5pt]
& BLBs  & BLBs   & BLBs  \\
Threshold & ($\log_{10}\tau=0$) & ($\log_{10}\tau=-0.8$) & ($\log_{10}\tau=-2.0$) \\
\midrule[1.5pt]
3.0\,kG & 100\%  & 100\%   & 87.8\% \\
3.5\,kG & 100\%  & 96.9\%  & 61.2\% \\
4.0\,kG & 100\%  & 84.7\%  & 31.6\% \\
4.5\,kG & 99.0\%  & 49.0\%  & 19.4\% \\
5.0\,kG & 93.9\% & 27.6\%  & 11.2\% \\
6.0\,kG & 68.4\% & 10.2\%  & 4.1\%  \\
\bottomrule[1.5pt]
\end{tabular}
    \caption{Percentages of BLBs with at least four adjacent pixels with magnetic field strengths larger than a threshold. Rows display thresholds ranging from 3 to 6\,kG. Columns 2--4 represent magnetic fields in the bottom node, middle node, and the top node, respectively. Percentages are given with respect to the 98 BLBs studied here.}
    \label{tab:fieldlargerthan}
\end{table}

Figure~\ref{fig:ObservedStokes} shows a few representative examples of BLBs. The first two columns display the continuum images of the entire Hinode/SOT-SP scan and the zoom into the region of interest (ROI). The BLBs are outlined by a green contour in the ROI (also in  Fig.~\ref{fig:FittedStokes}). The strong intensity gradient between the BLB and the umbra makes it easy to draw this contour line towards the adjacent umbrae on both sides. The end points of the BLB along the main axis connect smoothly to the adjacent penumbra, making an exact determination of their position challenging. Since the location of these end points is of low relevance for the analysis presented in this Letter, they were traced manually. The four rightmost columns of Fig.~\ref{fig:ObservedStokes} present the observed Stokes profiles along the spectrograph slit, which at least partly covers the BLB. All Stokes $I$ profiles show completely split Zeeman components at the location of the BLBs in the magnetically more sensitive \ion{Fe}{1} line at $6302.5$\,\AA{}.

Figure~\ref{fig:FittedStokes} shows the full Stokes vector for selected pixels arising within the BLBs (gray stars in second column). The polarisation signals are large, with several Stokes profiles exceeding the 10\% level. The presence of strong magnetic fields within BLBs is confirmed directly by the large Zeeman splitting of the Stokes profiles in 98.3\% of the \numblbsightings{} BLB scans. The blue lines show the fits achieved by the spatially coupled inversions, which reproduce the observations very well.

Ninety-seven out of 98 BLBs contain, in at least one Hinode/SOT-SP scan, magnetic fields stronger than 4.5\,kG at $\tau=1$.
For the middle and top nodes the percentage of fields stronger than 4.5\,kG decreases to  49\% (48/98)  and 19.4\% (19/98), respectively. Table~\ref{tab:fieldlargerthan} summarises the number of BLBs showing the percentages of BLBs harbouring field strengths above thresholds ranging from 3 to 6\,kG. Fig~\ref{fig:RF} in Appendix~\ref{sec:RF} presents the response functions (RFs) to a perturbation of $B$ as a function of optical depth.  The RF calculation reveals that the retrieval of $B$ at the three-node positions is robust.

The \modest{} inversions reveal that
the magnetic field strength is largest near the centres of the light bridges along their long axis, where they are furthest from the surrounding penumbra, at the sides adjacent to the two opposite-polarity umbrae. Fig.~\ref{fig:bones} shows the continuum images of extracted BLBs, where the location of strong magnetic fields larger than 3\,kG within BLBs are shaded red. 
We also checked the magnetic field strengths in the umbrae directly next to the locations of the strongest fields on the BLBs. In the places where a reliable determination of the umbral field is possible (i.e., where there are no or only weak molecular lines blending the Stokes signals), we confirm typical umbral values in the range of 2.5--4.0\,kG, clearly lower than the superstrong fields in the BLBs.

In Figs.~\ref{fig:AtmosphericConditionsBLBs}a--\ref{fig:AtmosphericConditionsBLBs}c, we compare the magnetic field strength at three optical depths against the normalised continuum intensity for umbrae (green), penumbrae (blue), and BLBs (red) for all \numblbsightings{} BLBs used in this study. The continuum intensity has been normalised such that it is unity for average quiet Sun in the surroundings of the sunspot.  The shaded areas represent  intensity histograms constructed from all individual spatial pixels of the corresponding feature within the Hinode/SOT-SP scan that hosts the BLB(s). Figure~\ref{fig:bones} illustrates that magnetic fields greater than 3 kG might not cover the entire mask used to isolate BLBs. To determine the strongest fields in each BLB, the green, blue, and red symbols are used to represent the average $B$ values of among the strongest 25\% of the pixels within the whole umbrae and penumbrae in the SOT-SP scan, and the BLBs, respectively. Averages taken with the strongest 50\% of pixels do not change the results. 

Figures~\ref{fig:AtmosphericConditionsBLBs}a--\ref{fig:AtmosphericConditionsBLBs}c show that these regions form three clearly separated populations, with almost no overlap. The mean magnetic field inside BLBs is systematically higher than in umbrae, with their continuum intensities being similar to those in penumbral regions. The magnetic field strength in the BLBs decreases significantly more rapidly with optical depth than the umbral and penumbral magnetic field strengths. The average vertical change of the magnetic field strength in BLBs is at least twice as large as the average vertical change in the penumbra and umbra (see Fig.~\ref{fig:gradient} and Tab.~\ref{tab:gradient}).

It is worth mentioning that the spectral lines of the  diatomic CaH and TiO molecules do not allow for a reliable magnetic field strength determination in the very cold umbrae \citep[see Fig~11 of][\citet{Berdyugina2011ASPC..mol6302}]{CastellanosDuran2024...modest}. For continuum intensities below 0.15, the umbral population is therefore likely to be incomplete and even stronger umbral magnetic fields remain undetected. However, these undetected strong fields would enhance the already present upward pointing tail of the umbral distribution in the deeper layers at very low intensity values, and therefore would clearly be separated from the population of superstrong fields in BLBs.

\section{Summary and discussion}

Regions within sunspot groups harbouring stronger magnetic fields than the umbra have so far been considered exceptional cases \citep[e.g.,][]{Tanaka1991SoPh, Okamoto2018ApJ, CastellanosDuran2020, Lozitsky2022ApJ...Strongfields}. We show here that the presence of extremely strong magnetic fields is rather common in  BLBs, i.e., most $\delta$-sunspots are expected to harbour very strong fields in the light bridges separating their opposite-polarity umbrae.

The presence of strong magnetic fields within BLBs can be directly measured by the Zeeman splitting of the observed Stokes profiles. The detailed atmospheric conditions within BLBs were obtained by applying spatially-coupled inversions assuming LTE conditions to the spectropolarimetric data, yielding excellent fits to the observed spectra (Fig.~\ref{fig:FittedStokes}). 

The superstrong magnetic fields in BLBs form a very distinct population in the magnetic field strength vs. continuum intensity diagram. The average magnetic field strengths in this population at the deepest observable layer is between 3 and 6\,kG, with individual pixels clearly exceeding this value. The magnetic field strength in the BLB population decreases significantly faster with height than for the umbral population. 

The high continuum intensities, and therefore the high temperatures of BLBs compared to the adjacent umbrae indicate that convection persists even in the presence of superstrong magnetic fields (Fig.~\ref{fig:AtmosphericConditionsBLBs}), indicating the presence of magnetoconvection processes in a new regime, namely very strong fields combined with magnetoconvection that is highly effective in transporting energy. BLBs thus appear to be the best pathway to probing this largely unexplored magnetoconvection regime \citep[see ][]{Hotta2020MNRAS...strongB}.

The frequent occurrence of superstrong fields in complex active regions may have implications for our understanding of solar active regions, possibly able to store larger amounts of magnetic energy than previously assumed.\\[0.2cm] 

{\small We thank the anonymous referee for  insightful comments that enhanced the clarity of our paper for a wider audience. This project has received funding from the European Research Council (ERC) under the European Union's Horizon 2020 research and innovation program (grant agreement No. 101097844 -- project WINSUN). Hinode is a Japanese mission developed and launched by ISAS/JAXA, with NAOJ as a domestic partner and NASA and UKSA as international partners. It is operated by these agencies in cooperation with ESA and NSC (Norway). }

\bibliographystyle{aasjournal}

\appendix

\begin{figure*}[htb]
\includegraphics[width=.98\textwidth]{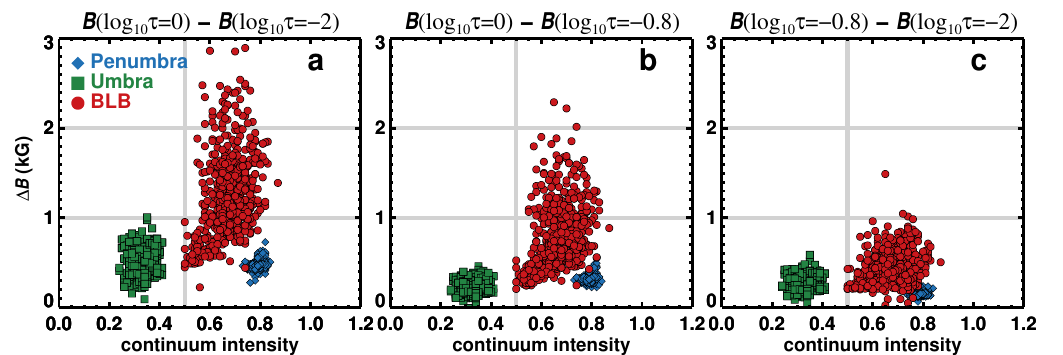}
\centering
 \caption{Vertical change of the magnetic field between nodes as a function of continuum intensity. Colours refer to penumbrae (blue), umbrae (green), and BLBs (red). Each BLB scan is represented by a symbol.}\label{fig:gradient}
 \end{figure*}

\begin{table*}[hpbt]
\centering
\begin{tabular}{cccc}
\toprule[1.5pt]
\multirow{2}{*}{Feature} & $B(\log_{10}\tau=0) - B(\log_{10}\tau=-2)$& $B(\log_{10}\tau=0) - B(\log_{10}\tau=-0.8)$ & $B(\log_{10}\tau=-0.8) - B(\log_{10}\tau=-2)$ \\
& (kG) & (kG) & (kG) \\
\midrule[1.5pt]
Penumbra & $0.48\pm0.05$ & $0.32\pm0.04$ & $0.16\pm0.03$ \\
Umbra & $0.5\pm0.1$ & $0.24\pm0.06$ & $0.28\pm0.08$ \\
BLB & $1.3\pm0.4$ & $0.8\pm0.3$ & $0.5\pm0.2$ \\
\bottomrule[1.5pt]
\end{tabular}
    \caption{Average change of the magnetic field between node positions depending on the solar feature within the sunspot group.}
    \label{tab:gradient}
\end{table*}
\section{Vertical change of the Magnetic field within bipolar light bridges}

BLBs exhibit a larger increase in magnetic field strength with optical depth than penumbrae or umbrae (see  Fig.~\ref{fig:AtmosphericConditionsBLBs}; the optical depth increases into the Sun). To determine whether this behaviour is a common feature of BLBs, we plot in  Fig.~\ref{fig:gradient} the vertical change of the magnetic field strength ($\Delta B$) for BLBs, penumbrae, and umbrae as a function of the continuum intensity. Each symbol represents  one of the 630 BLB scans, as well as the  umbrae and penumbrae associated with these scans. $\Delta B$ was estimated by subtracting the magnetic field strength at a lower node (larger optical depth) from that at a higher node. The three populations for the umbrae, penumbrae and BLBs are well separated. The average $\Delta B$ in BLBs is at least twice as large as the $\Delta B$ in the penumbrae and umbrae. However, for a few BLBs this is not the case.

Table~\ref{tab:gradient} summarizes the average differences for the three populations. The mean $\Delta B$ for the umbra and penumbra ranges between 0.16 and 0.5\,kG with a standard deviation up to 0.1\,kG. For BLBs, the mean $\Delta B$ is 1.3\,kG, and the scatter is more prominent with a standard deviation up to 0.5\, kG. 

\section{Examples of ARs with several bipolar light bridges}\label{sec:AR_many_BLBs}

\begin{figure*}[htbp]
\includegraphics[width=.98\textwidth]{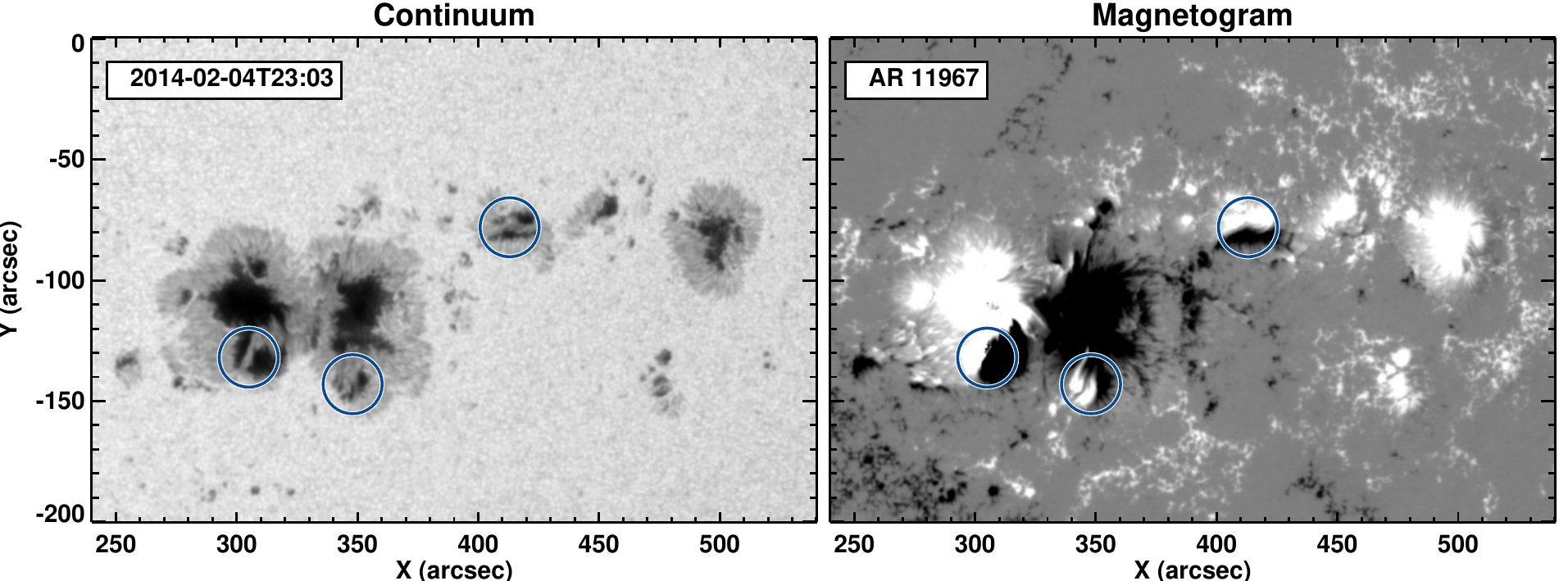}
\includegraphics[width=.98\textwidth]{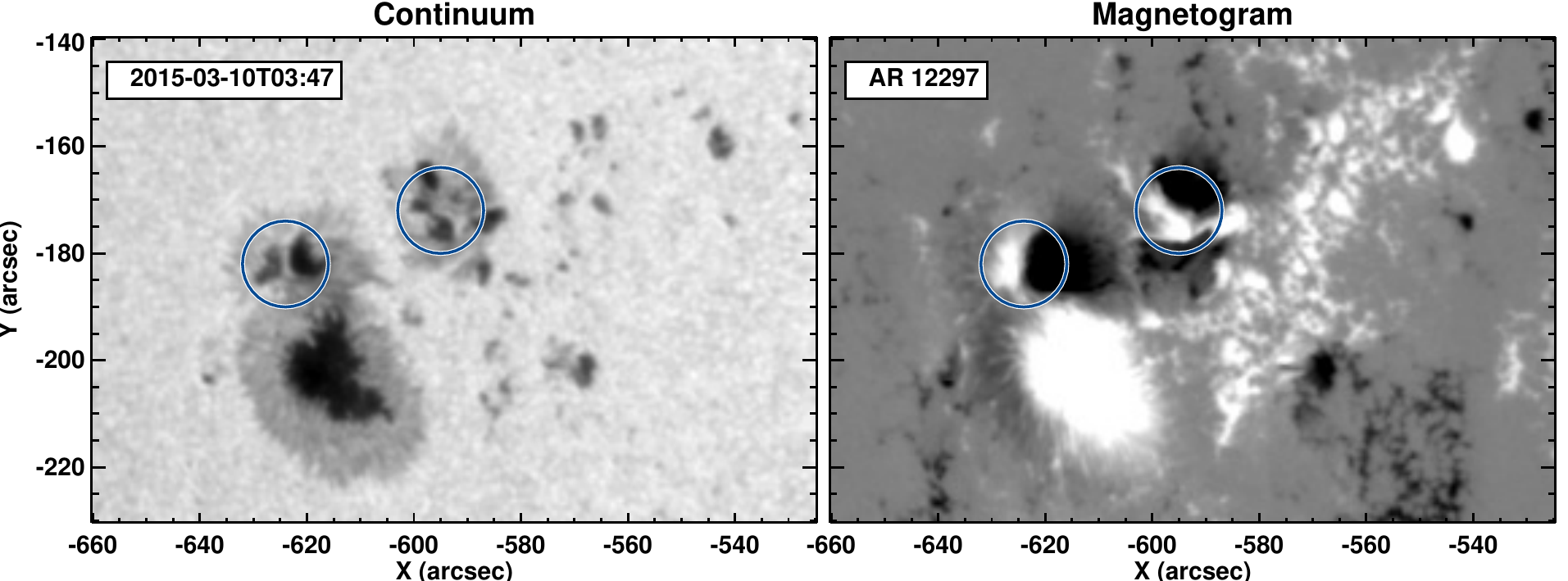}
\centering
 \caption{Continuum image (left) and magnetogram (right) of AR\,11967 (top) and AR\,12297 (bottom). Blue circles mark the locations of BLBs that appeared at the same time on different parts of this AR. }\label{fig:AR_with_many_BLBs}
 \end{figure*}

Multiple BLBs belonging to the same AR might appear at different phases of its evolution rather than necessarily all together. There are also complex ARs that host more exotic magnetic field configurations that harbour several BLBs at the same time. Typical examples are AR\,11967 (Fig.~\ref{fig:AR_with_many_BLBs}, top) and AR\,12297 (Fig.~\ref{fig:AR_with_many_BLBs}, bottom). Blue circles mark BLBs that appeared at the same time on different parts of these.

\section{Examples of the fragmentation and coalescence of bipolar light bridges}\label{sec:merging_fragmentation_BLBs}

The temporal evolution of BLBs shows that some fragment into two parts during their lifetime. Also, some BLBs merge as they evolve. In this appendix, we provide examples of BLBs that either fragmented or merged. 

\begin{figure}[htbp]
\includegraphics[height=.8\textheight]{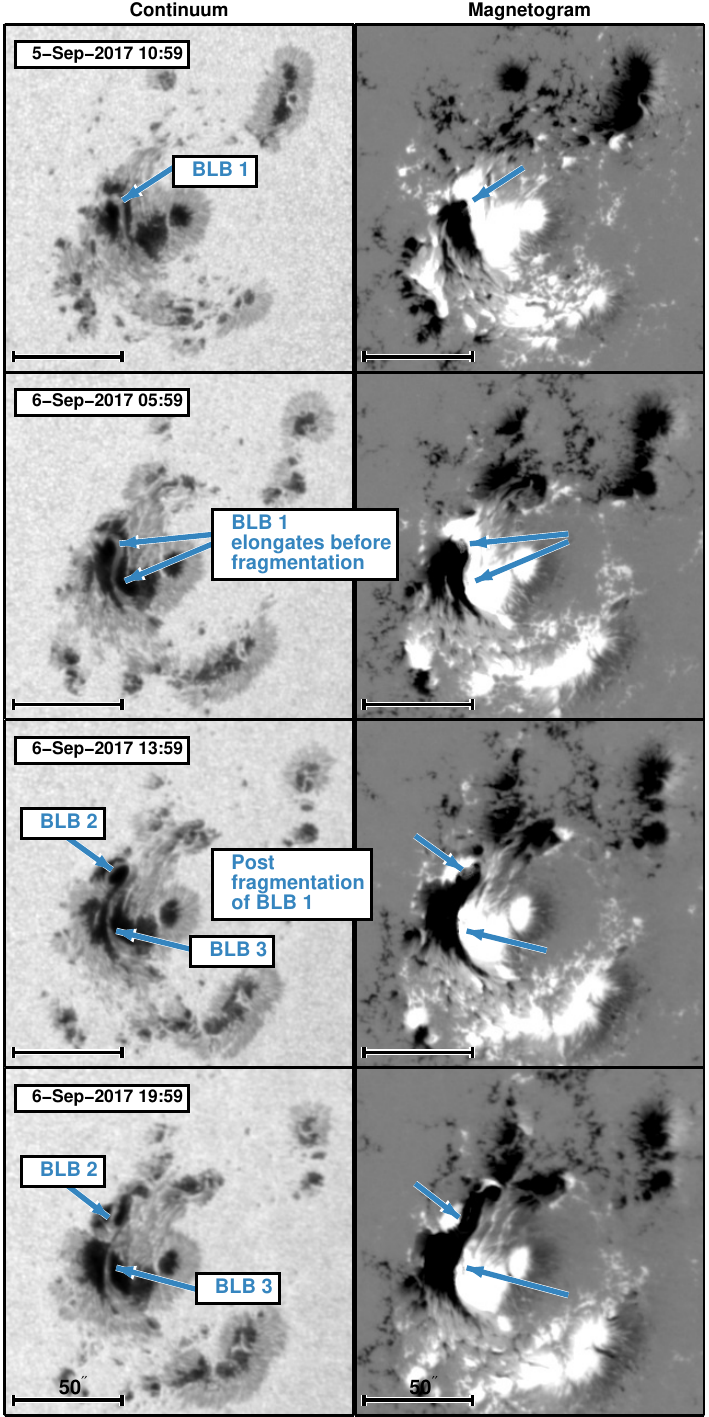}
\centering
 \caption{Fragmentation of BLB in AR\,12673 into two new BLBs. Continuum image (left) and magnetogram (right). Time runs from top to bottom. The bars denote 50\arcsec{}.}\label{fig:BLB_fragmentation}
 \end{figure}

Fragmentation mainly occurs in long BLBs, such as the one hosted by AR12673 (Fig.~\ref{fig:BLB_fragmentation}). The left panel of Fig.~\ref{fig:BLB_fragmentation} shows a long hooklike BLB. Then, the BLB elongates, stretches, and takes on a sigmoid shape (middle panel). Finally, the northern part of the BLB breaks off, producing a new delta spot. The fragmentation of the original BLB leads to the formation of two individual BLBs (right panel). In the temporal evolution, the connectivity of the LB changes to another region, but the two opposite polarities never coexist in direct contact.

\begin{figure}[htbp]
\includegraphics[height=.8\textheight]{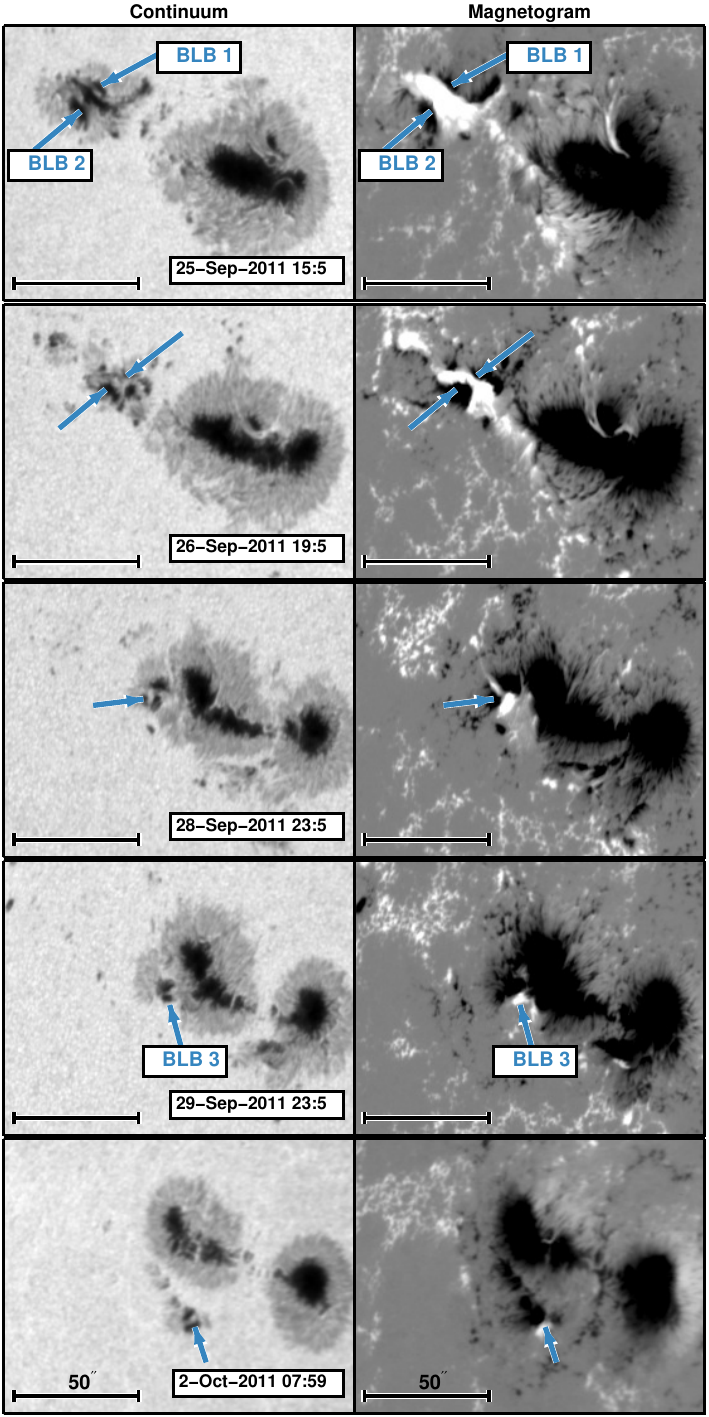}
\centering
 \caption{Coalescence of two BLBs in AR\,11302 into one BLB. The two BLBs are in the left spot of the delta group.  Continuum images (left) and magnetograms (right). Time runs from top to bottom. The bars denote 50\arcsec{}.}\label{fig:BLB_mergin}
 \end{figure}

Figure~\ref{fig:BLB_mergin} presents an example of two merging BLBs. A positive polarity umbra appears in between two negative polarity umbrae. Two BLBs form, one on either side of the positive polarity umbrae (top row). As time passes, the region's complexity decreases, and the two BLBs merge into one. The polarity of this feature is consistent throughout the whole passage of the AR on the disk. The images in Fig.~\ref{fig:BLB_mergin} show the transit of the AR from the eastern hemisphere to the western hemisphere. Again, the two opposite-polarity umbrae never coexist in direct contact. In both cases, merging or fragmentation, the opposite-polarity umbrae remain separated by a BLB.

\section{Response function to changes in $B$}\label{sec:RF}
\begin{figure*}[htbp]
\includegraphics[width=.99\textwidth]{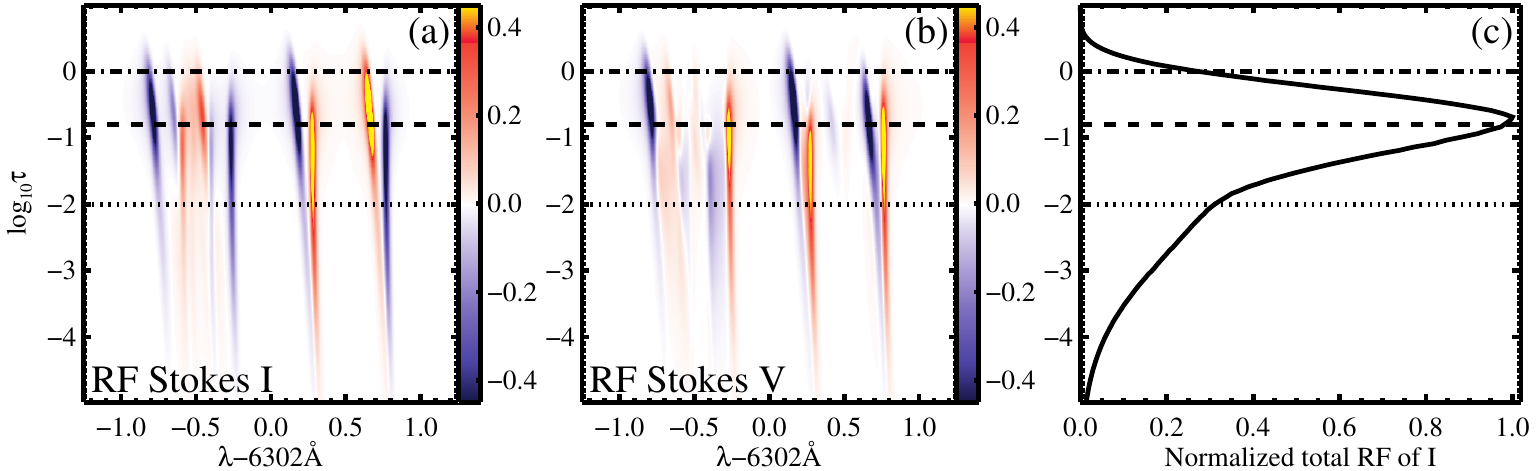}
\centering
 \caption{Normalized RFs for the two lines measured by Hinode/SP to changes in the magnetic field of a typical pixel within BLBs. Panels (a) and (b) show the RF of the Stokes profiles $I$ and $V$. Red (or blue) colours represent how the 6301.5\,\AA{} and 6302.5\AA{} Fe I lines are sensitive to a perturbation of $B$ at a given optical depth. Smaller optical depths correspond to greater heights in the solar atmosphere. Panel (c)  shows the total RF of Stokes $I$ as a function of optical depth. Vertical lines mark the optical depths at which the nodes of the \modest{} inversions are located.}\label{fig:RF}
 \end{figure*}

We calculated the response function (RF) to changes of $B$ of the Iron pair at 6301.5\,\AA{} and 6302.5\,\AA{} for a typical BLB pixel harbouring a magnetic field larger than 5\,kG (see Fig~\ref{fig:RF}). Panels (\ref{fig:RF}a) and (\ref{fig:RF}b) show the RFs of the Stokes profiles $I$ and $V$.  There is a non-vanishing response at optical depth equal to zero and even below. Panel (\ref{fig:RF}c) illustrates the contribution to the total RF of Stokes $I$, where the vertical lines mark the node positions of \modest{} inversions. The total RF of Stokes $I$ quickly rises between $\log_{10}\tau=0$ and -0.7. The middle node of our inversions is located close to the maximum contribution, and there is a significant contribution above $\log_{10}\tau=-2$. The computed RFs to the magnetic field confirm that the Stokes measurements are sensitive to the magnetic field at the location of all three nodes. The retrieved high values at $\log_{10}\tau=0$ are therefore not an error introduced by the extrapolation of the inversion code into a depth layer without response to magnetic fields, but likely to be real. Additional support comes from the fact that multiple such regions, with different-looking Stokes profiles, show similar strong magnetic fields over a spatially coherent region larger than the PSF of the telescope. 

Furthermore, if we assume that the field in the lower node would mainly be influenced by the field in the middle node, then the fact that it is almost always larger in the lowest node compared with the middle node would be difficult to explain. In this case, the value in the bottom node would fluctuate around that of the middle node and would not be consistently higher.

\bibliographystyle{aa}
{\tiny
\bibliography{references}}

\begin{thebibliography}{46}
\expandafter\ifx\csname natexlab\endcsname\relax\def\natexlab#1{#1}\fi

\bibitem[{{Berdyugina}(2011)}]{Berdyugina2011ASPC..mol6302}
{Berdyugina}, S.~V. 2011, in Astronomical Society of the Pacific Conference
  Series, Vol. 437, Solar Polarization 6, ed. J.~R. {Kuhn}, D.~M. {Harrington},
  H.~{Lin}, S.~V. {Berdyugina}, J.~{Trujillo-Bueno}, S.~L. {Keil}, \&
  T.~{Rimmele}, 219

\bibitem[{{Borrero} \& {Ichimoto}(2011)}]{Borrero2011LRSP}
{Borrero}, J.~M. \& {Ichimoto}, K. 2011, Living Reviews in Solar Physics, 8, 4

\bibitem[{{Castellanos~Dur\'an}(2022)}]{CastellanosDuran2022...Phd}
{Castellanos~Dur\'an}, J.~S. 2022, PhD thesis, University of G\"ottingen

\bibitem[{{Castellanos~Dur{\'a}n} {et~al.}(2023){Castellanos~Dur{\'a}n},
  {Korpi-Lagg}, \& {Solanki}}]{CastellanosDuran2023...ejectionCEFs}
{Castellanos~Dur{\'a}n}, J.~S., {Korpi-Lagg}, A., \& {Solanki}, S.~K. 2023,
  \apj, 952, 162

\bibitem[{{Castellanos~Dur{\'a}n} {et~al.}(2020){Castellanos~Dur{\'a}n},
  {Lagg}, {Solanki}, \& {Noort}}]{CastellanosDuran2020}
{Castellanos~Dur{\'a}n}, J.~S., {Lagg}, A., {Solanki}, S.~K., \& {Noort}, M.~v.
  2020, \apj, 895, 129

\bibitem[{{Castellanos Dur{\'a}n} {et~al.}(2024){Castellanos Dur{\'a}n},
  {Milanovic}, {Korpi-Lagg}, {L{\"o}ptien}, {van Noort}, \&
  {Solanki}}]{CastellanosDuran2024...modest}
{Castellanos Dur{\'a}n}, J.~S., {Milanovic}, N., {Korpi-Lagg}, A., {et~al.}
  2024, \aap, 687, A218

\bibitem[{{Collados} {et~al.}(1994){Collados}, {Martinez Pillet}, {Ruiz Cobo},
  {del Toro Iniesta}, \& {Vazquez}}]{Collados1994A&A...spots}
{Collados}, M., {Martinez Pillet}, V., {Ruiz Cobo}, B., {del Toro Iniesta},
  J.~C., \& {Vazquez}, M. 1994, \aap, 291, 622

\bibitem[{{Frutiger} {et~al.}(2000){Frutiger}, {Solanki}, {Fligge}, \&
  {Bruls}}]{Frutiger2000}
{Frutiger}, C., {Solanki}, S.~K., {Fligge}, M., \& {Bruls}, J.~H.~M.~J. 2000,
  \aap, 358, 1109

\bibitem[{{Hotta} \& {Toriumi}(2020)}]{Hotta2020MNRAS...strongB}
{Hotta}, H. \& {Toriumi}, S. 2020, \mnras, 498, 2925

\bibitem[{{Ichimoto} {et~al.}(2008){Ichimoto}, {Lites}, {Elmore}, {Suematsu},
  {Tsuneta}, {Katsukawa}, {Shimizu}, {Shine}, {Tarbell}, {Title}, {Kiyohara},
  {Shinoda}, {Card}, {Lecinski}, {Streander}, {Nakagiri}, {Miyashita},
  {Noguchi}, {Hoffmann}, \& {Cruz}}]{Ichimoto2008SoPh}
{Ichimoto}, K., {Lites}, B., {Elmore}, D., {et~al.} 2008, \solphys, 249, 233

\bibitem[{{Jur{\v{c}}{\'a}k} {et~al.}(2018){Jur{\v{c}}{\'a}k}, {Rezaei},
  {Gonz{\'a}lez}, {Schlichenmaier}, \&
  {Vomlel}}]{Jurcak2018A&A...JurcakCriterion}
{Jur{\v{c}}{\'a}k}, J., {Rezaei}, R., {Gonz{\'a}lez}, N.~B., {Schlichenmaier},
  R., \& {Vomlel}, J. 2018, \aap, 611, L4

\bibitem[{{Kiess} {et~al.}(2014){Kiess}, {Rezaei}, \&
  {Schmidt}}]{Kiess2014A&A...Bumbra}
{Kiess}, C., {Rezaei}, R., \& {Schmidt}, W. 2014, \aap, 565, A52

\bibitem[{{Kosugi} {et~al.}(2007){Kosugi}, {Matsuzaki}, {Sakao}, {Shimizu},
  {Sone}, {Tachikawa}, {Hashimoto}, {Minesugi}, {Ohnishi}, {Yamada}, {Tsuneta},
  {Hara}, {Ichimoto}, {Suematsu}, {Shimojo}, {Watanabe}, {Shimada}, {Davis},
  {Hill}, {Owens}, {Title}, {Culhane}, {Harra}, {Doschek}, \&
  {Golub}}]{Kosugi2007}
{Kosugi}, T., {Matsuzaki}, K., {Sakao}, T., {et~al.} 2007, \solphys, 243, 3

\bibitem[{{Li} {et~al.}(2022){Li}, {Zhang}, {Yan}, {Norton}, {Wang}, {Yang},
  {Xue}, \& {Kong}}]{Li2022ApJ...I2Binsunspots}
{Li}, Q., {Zhang}, L., {Yan}, X., {et~al.} 2022, \apj, 936, 37

\bibitem[{{Lites} \& {Ichimoto}(2013)}]{Lites2013SoPh}
{Lites}, B.~W. \& {Ichimoto}, K. 2013, \solphys, 283, 601

\bibitem[{{Liu} {et~al.}(2023){Liu}, {Sun}, {Schuck}, {Jaeggli}, {Welsch}, \&
  {Noda}}]{Liu2023ApJ...strongfieldandflows}
{Liu}, J., {Sun}, X., {Schuck}, P.~W., {et~al.} 2023, \apj, 955, 40

\bibitem[{{Livingston} {et~al.}(2006){Livingston}, {Harvey}, {Malanushenko}, \&
  {Webster}}]{Livingston2006SoPh}
{Livingston}, W., {Harvey}, J.~W., {Malanushenko}, O.~V., \& {Webster}, L.
  2006, \solphys, 239, 41

\bibitem[{{Lozitsky} {et~al.}(2022){Lozitsky}, {Yurchyshyn}, {Ahn}, \&
  {Wang}}]{Lozitsky2022ApJ...Strongfields}
{Lozitsky}, V., {Yurchyshyn}, V., {Ahn}, K., \& {Wang}, H. 2022, \apj, 928, 41

\bibitem[{{Martinez Pillet} \&
  {Vazquez}(1993)}]{MartinezPillet1993A&A...B2cont}
{Martinez Pillet}, V. \& {Vazquez}, M. 1993, \aap, 270, 494

\bibitem[{{Mathew} {et~al.}(2003){Mathew}, {Lagg}, {Solanki}, {Collados},
  {Borrero}, {Berdyugina}, {Krupp}, {Woch}, \&
  {Frutiger}}]{Mathew2003A&A...spot}
{Mathew}, S.~K., {Lagg}, A., {Solanki}, S.~K., {et~al.} 2003, \aap, 410, 695

\bibitem[{{Mathew} {et~al.}(2007){Mathew}, {Mart{\'\i}nez Pillet}, {Solanki},
  \& {Krivova}}]{Mathew2007A&A...spotcontrast}
{Mathew}, S.~K., {Mart{\'\i}nez Pillet}, V., {Solanki}, S.~K., \& {Krivova},
  N.~A. 2007, \aap, 465, 291

\bibitem[{{Okamoto} \& {Sakurai}(2018)}]{Okamoto2018ApJ}
{Okamoto}, T.~J. \& {Sakurai}, T. 2018, \apjl, 852, L16

\bibitem[{{Pevtsov} {et~al.}(2014){Pevtsov}, {Bertello}, {Tlatov}, {Kilcik},
  {Nagovitsyn}, \& {Cliver}}]{Pevtsov2014SoPh..CyclicB}
{Pevtsov}, A.~A., {Bertello}, L., {Tlatov}, A.~G., {et~al.} 2014, \solphys,
  289, 593

\bibitem[{{Pevtsov} {et~al.}(2011){Pevtsov}, {Nagovitsyn}, {Tlatov}, \&
  {Rybak}}]{Pevtsov2011ApJ...LongtermB}
{Pevtsov}, A.~A., {Nagovitsyn}, Y.~A., {Tlatov}, A.~G., \& {Rybak}, A.~L. 2011,
  \apjl, 742, L36

\bibitem[{{Schad}(2014)}]{Schad2014SoPh..UmbralB}
{Schad}, T.~A. 2014, \solphys, 289, 1477

\bibitem[{{Schad} \& {Penn}(2010)}]{Schad2010SoPh..UmbralB}
{Schad}, T.~A. \& {Penn}, M.~J. 2010, \solphys, 262, 19

\bibitem[{{Schuck} {et~al.}(2022){Schuck}, {Linton}, {Knizhnik}, \&
  {Leake}}]{Schuck2022ApJ...strongfields}
{Schuck}, P.~W., {Linton}, M.~G., {Knizhnik}, K.~J., \& {Leake}, J.~E. 2022,
  \apj, 936, 94

\bibitem[{{Siu-Tapia} {et~al.}(2017){Siu-Tapia}, {Lagg}, {Solanki}, {van
  Noort}, \& {Jur{\v c}{\'a}k}}]{Siu-Tapia2017A&A}
{Siu-Tapia}, A., {Lagg}, A., {Solanki}, S.~K., {van Noort}, M., \& {Jur{\v
  c}{\'a}k}, J. 2017, \aap, 607, A36

\bibitem[{{Siu-Tapia} {et~al.}(2019){Siu-Tapia}, {Lagg}, {van Noort}, {Rempel},
  \& {Solanki}}]{Siutapia2019}
{Siu-Tapia}, A., {Lagg}, A., {van Noort}, M., {Rempel}, M., \& {Solanki}, S.~K.
  2019, \aap, 631, A99

\bibitem[{{Smitha} {et~al.}(2020){Smitha}, {Holzreuter}, {van Noort}, \&
  {Solanki}}]{Smitha2020A&A......NLTEFeI}
{Smitha}, H.~N., {Holzreuter}, R., {van Noort}, M., \& {Solanki}, S.~K. 2020,
  \aap, 633, A157

\bibitem[{{Smitha} {et~al.}(2021){Smitha}, {Holzreuter}, {van Noort}, \&
  {Solanki}}]{Smitha2021A&A...NLTEFeI}
{Smitha}, H.~N., {Holzreuter}, R., {van Noort}, M., \& {Solanki}, S.~K. 2021,
  \aap, 647, A46

\bibitem[{{Smitha} {et~al.}(2023){Smitha}, {van Noort}, {Solanki}, \&
  {Castellanos~Dur{\'a}n}}]{Smitha2022....NLTE6173}
{Smitha}, H.~N., {van Noort}, M., {Solanki}, S.~K., \& {Castellanos~Dur{\'a}n},
  J.~S. 2023, \aap, 669, A144

\bibitem[{{Solanki}(1987)}]{Solanki1987PhDT}
{Solanki}, S.~K. 1987, PhD thesis, ETH, Z\"urich

\bibitem[{{Solanki}(2003)}]{Solanki2003}
{Solanki}, S.~K. 2003, \aapr, 11, 153

\bibitem[{{Sun} \& {Norton}(2017)}]{Sun2017RNAAS...strongfields}
{Sun}, X. \& {Norton}, A.~A. 2017, Research Notes of the American Astronomical
  Society, 1, 24

\bibitem[{{Tanaka}(1991)}]{Tanaka1991SoPh}
{Tanaka}, K. 1991, \solphys, 136, 133

\bibitem[{{Tiwari} {et~al.}(2015){Tiwari}, {van Noort}, {Solanki}, \&
  {Lagg}}]{Tiwari2015A&A...penumbra}
{Tiwari}, S.~K., {van Noort}, M., {Solanki}, S.~K., \& {Lagg}, A. 2015, \aap,
  583, A119

\bibitem[{Toriumi \& Hotta(2019)}]{Toriumi2019ApJL...strongB}
Toriumi, S. \& Hotta, H. 2019, \apj, 886, L21

\bibitem[{{Tsuneta} {et~al.}(2008){Tsuneta}, {Ichimoto}, {Katsukawa}, {Nagata},
  {Otsubo}, {Shimizu}, {Suematsu}, {Nakagiri}, {Noguchi}, {Tarbell}, {Title},
  {Shine}, {Rosenberg}, {Hoffmann}, {Jurcevich}, {Kushner}, {Levay}, {Lites},
  {Elmore}, {Matsushita}, {Kawaguchi}, {Saito}, {Mikami}, {Hill}, \&
  {Owens}}]{Tsuneta2008}
{Tsuneta}, S., {Ichimoto}, K., {Katsukawa}, Y., {et~al.} 2008, \solphys, 249,
  167

\bibitem[{{van Noort}(2012)}]{vanNoort2012A&A}
{van Noort}, M. 2012, \aap, 548, A5

\bibitem[{{van Noort} {et~al.}(2013){van Noort}, {Lagg}, {Tiwari}, \&
  {Solanki}}]{vanNoort2013A&A}
{van Noort}, M., {Lagg}, A., {Tiwari}, S.~K., \& {Solanki}, S.~K. 2013, \aap,
  557, A24

\bibitem[{{Verma}(2018)}]{Verma2018A&A}
{Verma}, M. 2018, \aap, 612, A101

\bibitem[{{Wang} {et~al.}(2018){Wang}, {Yurchyshyn}, {Liu}, {Ahn}, {Toriumi},
  \& {Cao}}]{Wang2018RNAAS}
{Wang}, H., {Yurchyshyn}, V., {Liu}, C., {et~al.} 2018, Research Notes of the
  American Astronomical Society, 2, 8

\bibitem[{{Yang} {et~al.}(2017){Yang}, {Zhang}, {Zhu}, \&
  {Song}}]{Yang2017ApJ...XflareBLB}
{Yang}, S., {Zhang}, J., {Zhu}, X., \& {Song}, Q. 2017, \apjl, 849, L21

\bibitem[{{Zirin} \& {Wang}(1993)}]{ZirinWang1993Natur...BipolarLightBridge}
{Zirin}, H. \& {Wang}, H. 1993, \nat, 363, 426

\bibitem[{Zirin \& Wang(1993)}]{Zirin1993b}
Zirin, H. \& Wang, H. 1993, \solphys, 144, 37

\end{thebibliography}

\end{document}